\documentclass[fleqn,usenatbib]{aa}
\usepackage[utf8]{inputenc}
\usepackage{comment}
\usepackage{orcidlink}

\usepackage{graphicx}
\usepackage{amsmath}
\usepackage{nccmath}
\usepackage{natbib} 
\usepackage{amssymb}
\usepackage{txfonts}
\usepackage{enumitem}
\usepackage[rightcaption]{sidecap}

\usepackage{newtxtext}

\usepackage[T1]{fontenc}

\DeclareRobustCommand{\VAN}[3]{#2}
\let\VANthebibliography\thebibliography
\def\thebibliography{\DeclareRobustCommand{\VAN}[3]{##3}\VANthebibliography}

\def \kms{km s$^{-1}$}

\defcitealias{fielding22}{FB22}

\newcommand{\hi}{\ifmmode{\rm HI}\else{H\/{\sc i}}\fi}

\usepackage{lmodern} 
\rmfamily 
\DeclareFontShape{T1}{lmr}{b}{sc}{<->ssub*cmr/bx/sc}{}
\DeclareFontShape{T1}{lmr}{bx}{sc}{<->ssub*cmr/bx/sc}{}

\usepackage{hyperref}
\hypersetup{
colorlinks=true,
citecolor=blue
}

\title{Modeling the Milky Way wind: Supernova-driven outflows\\ accelerate HI clouds near the Galactic center}

\titlerunning{Modeling the MW wind}

\author{
      Andrea Afruni\inst{1,2}\thanks{Email: andrea.afruni@unifi.it}\orcidlink{0000-0002-2858-6950}\
      \and
      Enrico M.~Di Teodoro\inst{1,2}\orcidlink{0000-0003-4019-0673}
      \and
      Lucia Armillotta\inst{1,2}\orcidlink{0000-0002-5708-1927} 
      \and \\
      Callum A. Lynn\inst{3}\orcidlink{0000-0001-6846-5347}
      \and
      N.~M.~McClure-Griffiths\inst{3}\orcidlink{0000-0003-2730-957X} 
      }
\authorrunning{A.~Afruni et al.}
     \institute{
     Dipartimento di Fisica e Astronomia, Universit\`a di Firenze, Via G. Sansone 1, 50019 Sesto Fiorentino, Firenze, Italy
     \and
     INAF - Osservatorio Astrofisico di Arcetri, Largo E. Fermi 5, Firenze, I-50125, Italy
    \and
    Research School of Astronomy and Astrophysics, The Australian National University, Canberra, ACT 2611, Australia
         }

\begin{document}

\abstract{
Multiwavelength observations, from radio to X-rays, have revealed the presence of multiphase high-velocity gas near the center of the Milky Way likely associated with powerful galactic outflows. This region offers a unique laboratory to study the physics of feedback and the nature of multiphase winds in detail. To this end, we have developed physically motivated semi-analytical models of a multiphase outflow consisting of a hot gas phase (\(T \gg 10^6\)~K) that embeds colder clouds (\(T \sim 5000\)~K). Our models include the gravitational potential of the Milky Way; the drag force exerted by the hot phase onto the cold clouds; and the exchange of mass, momentum, and energy between gas phases.
Using Bayesian inference, we compared the predictions of our models with observations of a population of \hi\ high-velocity clouds detected up to $\sim$1.5~kpc above the Galactic plane near the Galactic center. We find that a class of supernova-driven winds launched by star formation in the central molecular zone can successfully reproduce the observed velocities, spatial distribution, and masses of the clouds. In our two-phase models, the mass and energy loading factors of both phases are consistent with recent theoretical expectations. The cold clouds are accelerated by the hot wind via ram pressure drag and via accretion of high-velocity material, resulting from the turbulent mixing and subsequent cooling. However, this interaction also leads to gradual cloud disruption, with smaller clouds losing over 70\% of their initial mass by the time they reach $\sim$2~kpc.
}


\maketitle

\section{Introduction}\label{intro}
Galaxies eject material into their surroundings through powerful gas outflows, which can be powered by processes related either to star formation, such as supernovae (SNe) explosions and stellar winds, or to active galactic nuclei (AGNe). Multiphase outflows, from ionized to neutral and molecular gas phases have been extensively detected both in the local and in the distant Universe \citep[see][for a comprehensive review]{veilleux20}.  
From a theoretical point of view, a thermally driven hot wind that can accelerate cold material away from the galactic disk has been described by analytical and semi-analytical models with various levels of complexity, both for SN- and AGN-driven winds \citep[e.g.,][]{chevalier85,king15,fielding22,thompson24}. The multiphase nature of outflows has also emerged from high-resolution hydrodynamical simulations of galactic winds \citep[e.g.,][]{Gatto17,kim18,kim20, rathjen21}, where typically most of the energy is in the hot ($T \gg 10^6$ K) phase and most of the mass (at least at the wind base) is in the cold ($T\lesssim10^4$ K) phase \citep[see, e.g.,][]{li20}, which may eventually return to the disk or evaporate in the hot wind \citep[e.g.,][]{schneider20}. However, due to the complexity of the processes at play, to the limitations of the observational data, and to the wide range of scales involved, detailed comparisons between theoretical models and observations are still missing.

A unique laboratory to observe and study multiphase galactic outflows is the Milky Way (MW) galactic center. The detection of the $\gamma$-ray emitting structures extending up to latitudes of about $55^{\circ}$, called Fermi Bubbles \citep[][]{su10fb}, and more recently of their X-ray counterparts \citep[eRosita Bubbles, ][]{predehl20}, are a clear sign of an energetic outflow event from the central region of our Galaxy. Despite this evidence, whether such an event is due to the activity of the supermassive black hole (SMBH) at the center of the MW \citep[e.g.,][]{guo12} or to central star formation activity \citep[e.g.,][]{crocker15, nguyen22}, which is particularly high ($\sim0.1\ M_{\odot}/\rm{yr}$) in a central ring with a radius of a few hundred parsecs known as the central molecular zone \citep[CMZ; see][]{henshaw23}, is still a matter of debate \citep[see review by][and references therein]{sarkar24}. Warm and cold gas exhibiting kinematics consistent with an outflow have also been detected in UV absorption \citep[e.g.,][]{fox15,bordoloi17,cashman23}, neutral hydrogen (\hi) line emission at 21 cm \citep{cluregriffiths13,diteodoro18}, and down to molecular gas emission \citep[][]{diteodoro20,Noon23,heyer25} and absorption \citep{Cashman+21}. 

\begin{figure*}
    \centering
    \includegraphics[clip, trim={0cm 0cm 0cm 0cm}, width=\linewidth]{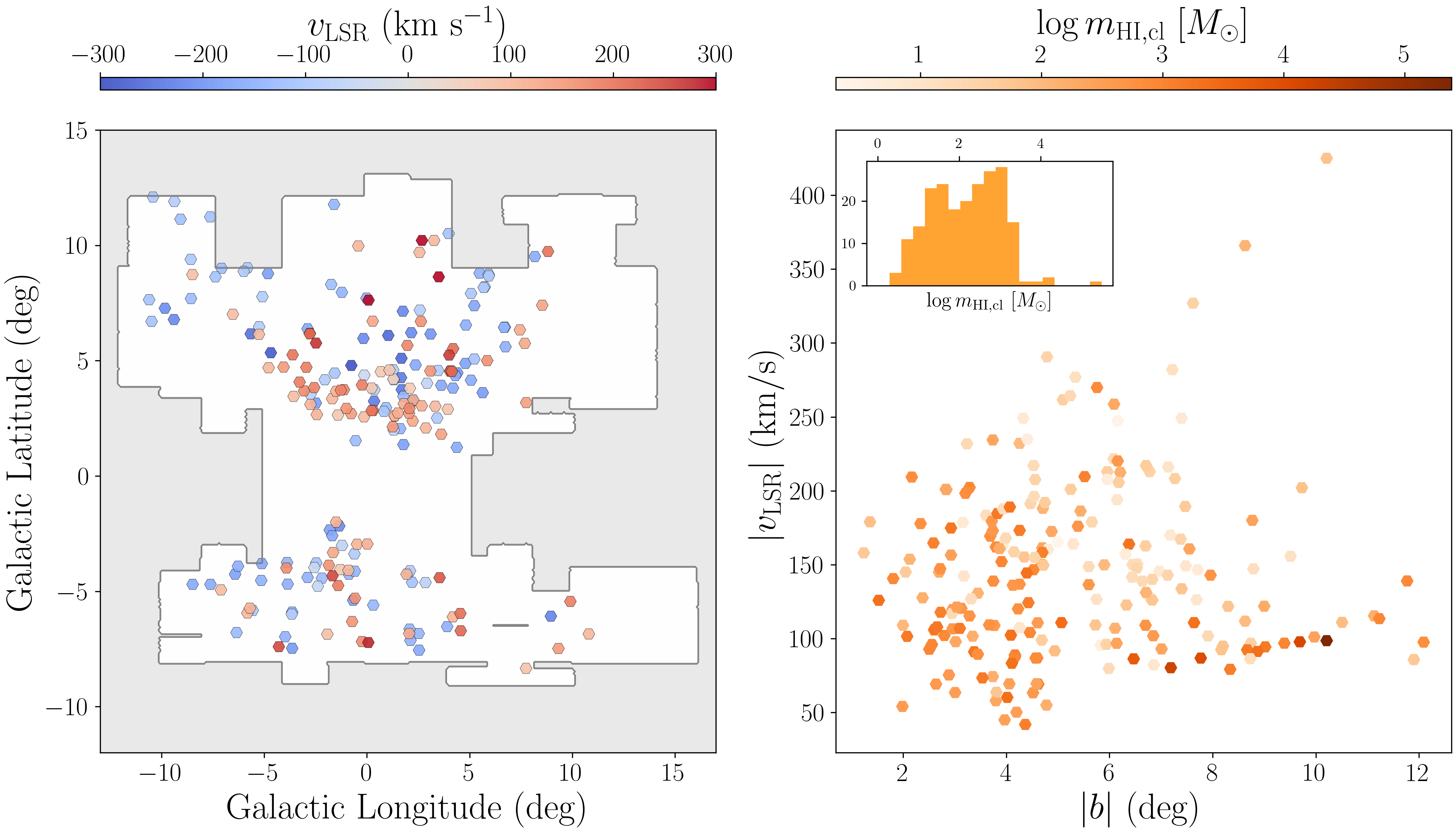}
    \caption{Population of \hi\ high-velocity clouds from the surveys of \cite{cluregriffiths13} and \cite{diteodoro18}, taken with the ATCA and the GBT. Left: Galactic latitude-longitude map. The data-points show the positions of the clouds and are color-coded by their $v_{\rm{LSR}}$. The white mask shows the region surveyed by the ATCA and the GBT. Right: \hi\ cloud latitudes as a function of their velocities. The kinematic pattern shows a signature of acceleration (see main text and \citealt{lockman20}). The points are color-coded by the cloud mass and the full 1D mass distribution is shown in the inset panel.}
    \label{fig:data}
\end{figure*} 

The plethora of multiwavelength data of the Galactic center outflow gives us the opportunity to study in detail the interplay between the different gas phases of a wind. In this work, we focus on the cool phase seen in 21-cm emission close to the center of the MW (see Section~\ref{observations}), which accurately traces the gas kinematics and therefore can be used to test theoretical models of the Galactic wind dynamics. In recent years, these data have been interpreted using simple kinematic models of a cold outflow that has a constant velocity or accelerates with latitude \citep[see][]{diteodoro18,lockman20}. Our main goal is to follow up on such models, developing a more physically motivated scenario where a hot, fast wind powered by SN feedback entrains the cold clouds. While the physical properties and dynamics of the \hi\ clouds close to the Galactic center have recently been studied using idealized high-resolution hydrodynamical simulations \citep[][]{zhang2024surv,zhang25acc}, in this paper we adopt semi-analytical models, with free parameters that we tune in order to reproduce the observed cloud kinematics. Although clearly simplified compared to simulations, this approach is more flexible and allows for comparison of observational data with model predictions that depend on key physical parameters. 

The remainder of this paper is structured as follows. In Section~\ref{observations} we summarize the \hi\ surveys that we use in this work, and we report the main observational features that we aim to reproduce with our models. In Section~\ref{modelSection}, we describe the details of our multiphase wind models. In Section~\ref{ModelResults}, we report the main findings of this work, which we obtained by comparing models and data through a Bayesian analysis. Finally, in Section~\ref{discussionSection}, we discuss the comparison of our best-fit models with additional observational data and the main uncertainties of our framework, while in Section~\ref{conclusions} we summarize our work, and we outline our conclusions.

\section{HI observations}\label{observations}
The results of this work are based on two \hi\ emission surveys of a region close to the Galactic center, presented in \cite{cluregriffiths13} and \cite{diteodoro18} and taken respectively with the Australia Telescope Compact Array (ATCA) and with the Green Bank Telescope (GBT). Combined, the two surveys span a region within Galactic latitudes of $|b|\lesssim 10^{\circ}$, with the first reaching up to $|b|\sim 5^{\circ}$ and the second focused instead at $5^{\circ}\lesssim|b|\lesssim10^{\circ}$. The exact extent of the surveyed region is shown by the white mask in the left panel of Figure~\ref{fig:data}. 

The main finding of these surveys is the discovery of a population of high-velocity \hi\ clouds, with local standard of rest (LSR) velocities, $v_{\rm{LSR}}$, up to $\sim400$ \kms\ and a kinematic pattern that is inconsistent with the Galactic rotation and likely associated with a Galactic nuclear outflow \citep[see][]{diteodoro18}. The total sample consists of 213 different components.\footnote{Compared to the catalogs presented in \citet{cluregriffiths13} and \citet{diteodoro18}, we included 21 additional clouds that were not present in the original samples. A more thorough analysis of the observational properties of the full sample will be presented in Yu et al. (in preparation).} In the left-hand side of Figure~\ref{fig:data} we report the positions of the 213 clouds in Galactic longitude ($l$) and latitude, with the color scale showing the cloud $v_{\rm{LSR}}$. Note that the absence of clouds within $|b|<2^{\circ}$ is mainly due to the confusion of the velocity components with the emission of the \hi\ Galactic disk. In addition, clouds with $v_{\rm{LSR}}<50$ \kms\ for the ATCA sample and with $v_{\rm{LSR}}<75$ \kms\ for the GBT sample in general cannot be easily distinguished from the \hi\ emission from the local MW disk and are currently excluded from the catalogs. A more systematic search for clouds with low $v_{\rm{LSR}}$ will be presented in Yu et al. (in preparation).

The most evident property of the \hi\ cloud kinematics is shown in the right-hand side of Figure~\ref{fig:data}: there is a clear sign of acceleration with increasing latitudes, as clouds closer to the disk tend to have lower LSR velocities. This feature was already identified by \cite{lockman20}, who explained it with an outflow model that accelerates from a velocity of about 200 \kms\ in the inner regions to a velocity of 330 \kms\ at a distance of $2.5$ kpc from the center. The aim of the current work is to expand on this finding by creating physical models of a multiphase wind where the cold clouds are accelerated by the fast-moving hot phase.

Additional properties of both samples, including cloud \hi\ column densities, masses, and radii, are reported in  \citet{cluregriffiths13} and \cite{diteodoro18}. The clouds have typical column densities of up to a few times $10^{19}\ \rm{cm}^{-2}$ and radii that go from a few parsecs to about $50$ pc. In the right panel of Figure~\ref{fig:data} we have color coded the data points with the estimated cloud masses. While there does not seem to be a significant trend of the mass with cloud velocity or distance from the center, we note that the majority of the clouds with $v_{\rm{LSR}}\gtrsim250$ \kms\ have relatively low masses ($m_{\rm HI,cl}\lesssim10^2\ M_{\odot}$). We also report in the inset in the same panel the one-dimensional mass distribution (the values have been updated with respect to those reported in \citealt{diteodoro18}). The clouds have masses that range from a few solar masses up to $10^5\ M_{\odot}$. We used the same mass distribution in our modeling (see Section~\ref{sec:modelCold}).

\section{Modeling the wind}\label{modelSection}
In order to interpret the data reported in Section~\ref{observations}, we developed semi-analytical models of a multiphase wind of the MW, with the main assumption that the outflow is powered by SN feedback due to the intense star formation in the CMZ (but see also Section~\ref{disc:WindAssumptions}).\footnote{In Appendix~\ref{sec:ballistic}, we also describe a single-phase ballistic model and we show that it cannot reproduce the acceleration feature seen in the \hi\ data.} The idea is based on the steady-state wind model of \cite{chevalier85}: these authors have shown that solving the hydrodynamic equations of the conservation of mass, momentum, and energy assuming constant injection rates of mass $\dot{M}_{\rm{inj}}$ and energy $\dot{E}_{\rm{inj}}$ within an inner radius $r_{\rm{inj}}$ leads to a fast-moving, high-temperature wind that quickly becomes supersonic outside of $r_{\rm{inj}}$. The hot wind in turn can entrain and accelerate the \hi\ cold clouds, potentially leading to the kinematic pattern observed through 21-cm data. We built our model on the more sophisticated framework developed by \cite{fielding22}, hereafter \citetalias{fielding22}, which describes the steady-state structure of a multiphase galactic wind, with the cloud-wind interaction parametrized based on recent wind-tunnel simulations. The details of our modeling are described below.

\subsection{Hot phase}\label{sec:modelHot} 
With respect to the classical model of \cite{chevalier85}, \citetalias{fielding22} considered the effects of the galaxy gravitational potential $\Phi$ and of the radiative cooling and heating $\mathcal{L}$ on the hot SN-driven wind. They then assumed that the spherically symmetric wind subtends a solid angle $\Omega$, which in our study we assume being equal to the full $4\pi$. Assuming a smaller solid angle, and hence a biconical wind, would only change the hot gas density and pressure normalization by an order-unity factor and would therefore not affect our final findings (see also Section~\ref{disc:WindAssumptions}). Following \citetalias{fielding22}, the three conservation equations for the hot gas can be rewritten, taking into account the physical assumptions outlined above, to describe more explicitly the evolution of the hot wind velocity, $v$; density, $\rho$; and pressure, $P$, as a function of the spherical radius, $r$:

\begin{ceqn}
\begin{align}
\frac{\partial \log v}{\partial \log r} &= \frac{1}{1 - \mathcal{M}^{-2}} \left( \frac{2}{\mathcal{M}^2} - \frac{v_c^2}{v^2} - \frac{\dot{\rho}_{\rm{inj}}/\rho}{v/r} \left( \frac{\gamma + 1}{2} \right) \right. \nonumber \\
&\quad \left. - \gamma \frac{\dot{p}_{\rm{inj}}}{\rho v} + (\gamma - 1) \frac{\dot{\mathcal{E}}_{\rm{inj}} - \mathcal{L}}{\rho v^2} + \frac{\gamma - 1}{2} \frac{v_{\text{esc}}^2}{v^2} \right), \\ 
\frac{\partial \log \rho}{\partial \log r} &= \frac{1}{1 - \mathcal{M}^{-2}} \left( -2 + \frac{v_c^2}{v^2} + \frac{\dot{\rho}_{\rm{inj}}/\rho}{v/r} \left( \frac{\gamma + 3}{2} \right) \right. \nonumber \\
&\quad \left. - \gamma \frac{\dot{p}_{\rm{inj}}}{\rho v} + (\gamma - 1) \frac{\dot{\mathcal{E}}_{\rm{inj}} - \mathcal{L}}{\rho v^2} - \frac{1}{\mathcal{M}^2} + \frac{\gamma - 1}{2} \frac{v_{\text{esc}}^2}{v^2} \right), \\ 
\frac{\partial \log P}{\partial \log r} &= \frac{\gamma}{1 - \mathcal{M}^{-2}} \left( -2 + \frac{v_c^2}{v^2} \right. \nonumber \\
&\quad + \frac{\dot{\rho}_{\rm{inj}}/\rho}{v/r} \left(1 + \frac{\gamma - 1}{2} \mathcal{M}^2 \left( 1 + \frac{v_{\text{esc}}^2}{v^2} \right) \right) - \frac{\dot{p}_{\rm{inj}}}{\rho v} \nonumber \\
&\quad \left. + ((\gamma - 1) \mathcal{M}^2) \frac{\dot{\mathcal{E}}_{\rm{inj}} - \mathcal{L} - \dot{p} v}{\rho v^2} \right).
\end{align}
\end{ceqn} 
Here $\mathcal{M}=v/c_{\rm{s}}$ is the Mach number, with $c^2_{\rm{s}}=\gamma P/\rho$ being the sound speed and $\gamma=5/3$ the adiabatic index; $v^2_{\rm{c}}=r\,d\Phi/dr$ is the circular velocity; and  $v^2_{\rm{esc}}=-2\Phi$ is the escape velocity. 
In this model, within the injection radius $r_{\rm{inj}}$, we have a constant mass density injection rate $\dot{\rho}_{\rm{inj}}=\dot{M}_{\rm{inj}}/(4\pi r^3_{\rm{inj}}/3)$ and an energy density injection rate in $\dot{\mathcal{E}}_{\rm{inj}}=\dot{E}_{\rm{inj}}/(4\pi r^3_{\rm{inj}}/3)$, with $\dot{M}_{\rm{inj}}$ and $\dot{E}_{\rm{inj}}$ being the mass and energy injection rate, respectively. 
For $r>r_\mathrm{inj}$, $\dot{\rho}_{\rm{inj}}=\dot{\mathcal{E}}_{\rm{inj}}=0$. We also assumed no momentum injection, i.e., $\dot{p}_{\rm{inj}}=0$ everywhere (as in the original model from \citealt{chevalier85}).

Throughout this work, we assumed that the hot gas is subject to a logarithmic gravitational potential with a circular velocity $v_{\rm{c}}=150$ \kms. As we demonstrate later, this is a simplification that is slightly inconsistent with our treatment of the dynamics of the cold clouds, but the gravitational potential only marginally influences the properties of the hot wind (if the wind densities are not too high, which is the case for our best-fit models, see Figure~\ref{fig:HotgasProp}) within 2 kpc from the disk (our region of interest), so our results do not depend on this choice. As for the cooling, we used the fiducial cooling tables used by \citetalias{fielding22}, who adopt the solar-metallicity cooling curves from \cite{wiersma09}, which also include the effects of the extragalactic UV background at $z=0$. We checked that a variation in the metallicity of the hot (and/or cold, see below) gas within a factor of 2 would not significantly affect the wind properties, and hence our final findings.

Finally, the injections of mass and energy within $r_{\rm{inj}}$ are related to the star formation rate, $SFR$, through the so-called mass and energy loading factors, $\eta_{\rm{M}}$ and $\eta_{\rm{E}}$, such that $\dot{M}_{\rm{inj}}=\eta_{\rm{M}}SFR$ and $\dot{E}_{\rm{inj}}=\eta_{\rm{E}}E_{\rm{SN}} SFR/(100\ M_{\odot})$, where $E_{\rm{SN}}=10^{51}\ \rm{erg}$ is the energy deposited by one supernova explosion and assuming that one supernova occurs every 100 $M_{\odot}$ formed. We assumed $SFR=0.1\ M_{\odot}/\rm{yr}$, which is consistent with current observational estimates of the $SFR$ in the CMZ (see Section~\ref{disc:WindAssumptions}). The injection radius and the two loading factors are the first three free parameters of our model (see Table~\ref{tab:parameters}, where we also summarize our choices for the fixed parameters described above).

\begin{figure*}
    \centering
    \includegraphics[clip, trim={0cm 0cm 0cm 0cm}, width=\linewidth]{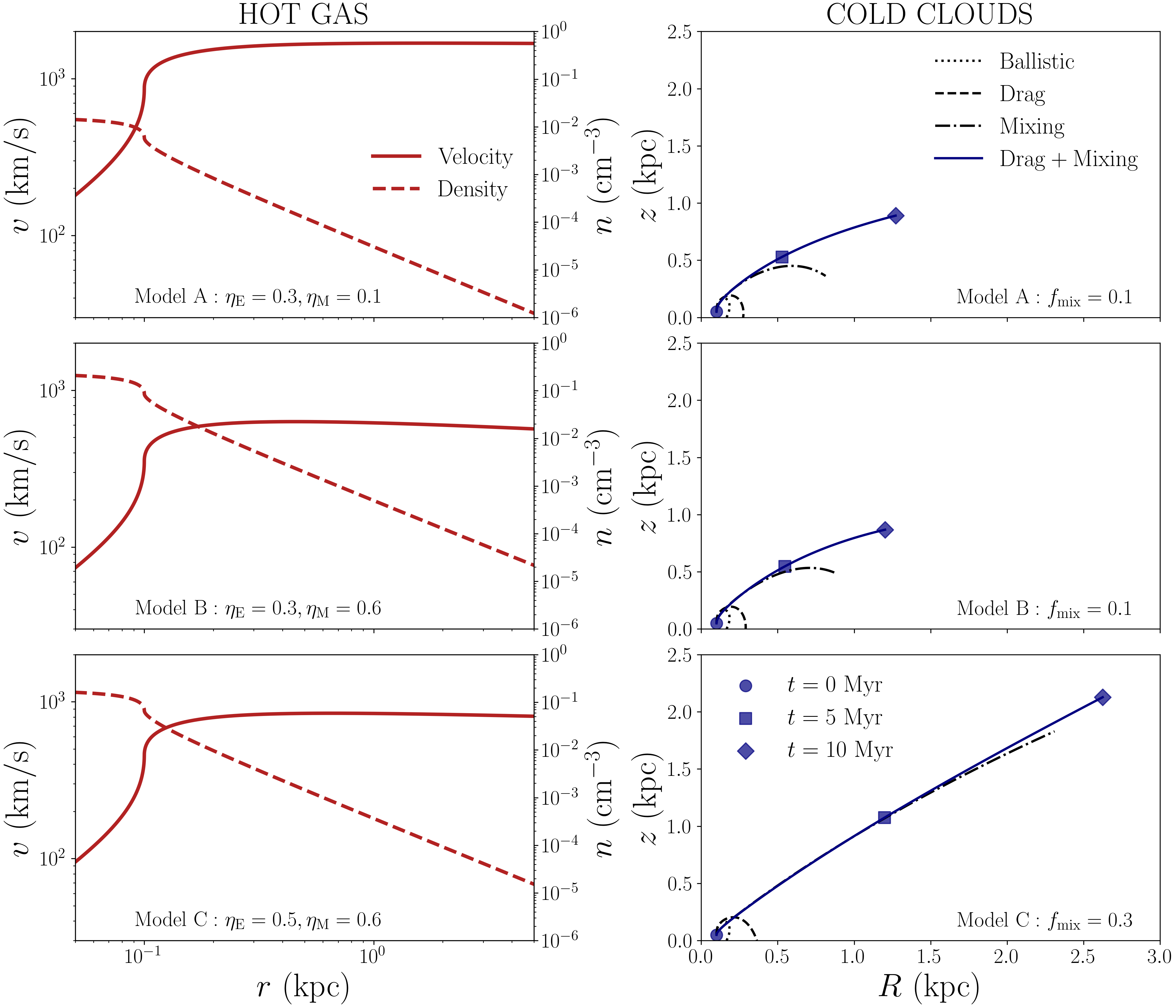}
    \caption{Behavior of the multiphase wind for three different choices of models (see main text): model A (top row), model B (central row), and model C (bottom row). On the left, hot gas velocities (solid curves) and densities (dashed curves) are shown as a function of the distance, $r$. The right panels show instead the cold cloud orbits (solid curves), indicating in particular the effects of drag (dashed curves) and mixing (dash-dotted curves) on the original ballistic orbit (dotted curves). The symbols mark the positions of the modeled cloud at the beginning (circles), after 5 Myr (squares) and after 10 Myr (rhomboids).}
    \label{fig:meth_orb}
\end{figure*}

In the left panels of Figure~\ref{fig:meth_orb} we show the velocity (solid curves) and density (dashed curves) profiles of three different models where we assumed the same injection radius, $r_{\rm{inj}}=100$ pc, and three different choices for the mass and energy loading factors: $\eta_{E}=0.3$, $\eta_{\rm{M}}=0.1$ (model A, top row); $\eta_{E}=0.3$, $\eta_{\rm{M}}=0.6$ (model B, central row); and $\eta_{E}=0.5$, $\eta_{\rm{M}}=0.6$ (model C, bottom row). One can see how in all cases the wind accelerates within $r_{\rm{inj}}$ to then reach a terminal supersonic velocity, 
while its density decreases substantially as it expands. As expected, increasing the mass loading factor implies that the wind becomes more massive and slower (the density increases by about an order of magnitude from model A to model B, while the terminal velocity goes from more than 1000 \kms\ to about 600 \kms), while an increase in the energy loading factor increases the wind velocity, without significantly affecting the density profile (see the differences between model B and C). The hot gas profiles of Figure~\ref{fig:meth_orb} are in line with the predictions of the more simplistic model of \cite{chevalier85}, meaning that both the gravitational potential and the radiative losses have a minimal effect on the wind properties (especially for models A and C). In the next section we discuss how different hot gas models affect differently the properties and the dynamics of the cold \hi\ clouds.

\subsection{Cold cloud dynamics}
\label{sec:modelCold}
The main goal of this work is to reproduce the properties and especially the kinematics of the \hi\ clouds presented in Section~\ref{observations}. For this purpose, we modeled the orbits of a population of clouds using a modified version of the python package {\sc galpy} \citep{bovy15}, which includes not only the effects of the gravitational potential, but also of the interactions between the clouds and the hot gas phase described above, in a similar fashion to the works of \cite{afruni21,afruni22} and following the framework of \citetalias{fielding22}. We assumed that the clouds have solar metallicity;\footnote{This is chosen for simplicity to be the same as the hot wind, given that we expect both phases to have $Z\gtrsim Z_{\odot}$ and that the exact metallicity values are highly uncertain from observations. A small metallicity gradient would anyway not significantly impact our findings.} a temperature $T_{\rm{cl}}=5000\ \rm{K}$ \citep[see][]{diteodoro18}; and that they are continuously pressure-confined by the hot gas, such that $\rho_{\rm{cl}}=\mu m_{\rm{p}}P/k_{\rm{B}}T_{\rm{cl}}$, with $\mu=0.62$. We note that this value of the mean molecular weight is appropriate only for the hot ionized gas, while for the cold phase it is $\mu_{\rm{cold}}\approx1.2$: we adopted this simplification following \citetalias{fielding22}, but a more accurate mean molecular weight should have a negligible impact on our findings.

\subsubsection{Gravitational potential and multiphase interactions}
We adopted one of the MW potentials available in {\sc galpy}, the \texttt{MWPotential2014}. While we refer the reader to \cite{bovy15} for additional details, in brief this potential has a circular velocity of $220$ \kms\ at $8$ kpc from the center and it is composed of three components: i) a spherical bulge that is modeled as a power-law density profile with an exponential cut-off; ii) a disk that is described by a Miyamoto-Nagai potential \citep{Miyamoto+1975}; and iii) a dark matter halo that is described by a Navarro-Frenk-White \citep[NFW,][]{nfw96} profile. While this is not the most recent or accurate representation of the potential of our Galaxy, it is the most convenient for the purpose of this work as its simplicity significantly increases the computational speed of the final model with only marginal differences compared to a more sophisticated potential, especially for orbits in the inner Galaxy.

While the gravitational pull of the MW tends to decelerate the clouds ejected from the disk, the hot wind has the opposite effect of entraining and accelerating the clouds compared to a simple ballistic model (see Appendix~\ref{sec:ballistic}). We treated this entrainment by following the framework of \citetalias{fielding22}, where the acceleration of a single spherical cloud of mass $m_{\rm{cl}}$ is driven, once the gravitational force is taken into account, by two additional terms. The first one is due to the drag force exerted by the hot gas \citep[see also e.g.,][]{fraternali08,marinacci11}:

\begin{ceqn}
\begin{equation}\label{eq:drag}
\dot{v}_{\rm{cl,drag}} = \frac{1}{2} C_{\rm{drag}}\rho(v-v_{\rm{cl}})^2\frac{A_{\rm{cross}}}{m_{\rm{cl}}}\ ,
\end{equation}
\end{ceqn}
where $A_{\rm{cross}}=\pi r^2_{\rm{cl}}$ is the cloud cross section, with $r_{\rm{cl}}=[3m_{\rm{cl}}/(4\pi\rho_{\rm{cl}})]^{1/3}$. We assume $C_{\rm{drag}}=1/2$ for consistency with \citetalias{fielding22}, who adopted this choice as it is applicable to flows with high Reynolds numbers.
 
The second term is instead due to hydrodynamical instabilities that cause mixing at the cloud interface between the low-velocity cold gas and the high-velocity hot gas: if the cooling time is sufficiently short, the mixed gas cools down and increases the mass of the cloud at a rate $\dot{m}_{\rm{cl,growth}}$. This mass transfer from the hot wind into the cloud leads then to a transfer of momentum that significantly accelerates the cloud, so that (see \citetalias{fielding22})
\begin{ceqn}
\begin{equation}\label{eq:acc_growth}
\dot{v}_{\rm{cl,growth}} = (v-v_{\rm{cl}}) \frac{\dot{m}_{\rm{cl,growth}}}{m_{\rm{cl}}}\ .
\end{equation}
\end{ceqn}
For an in-depth discussion on how the cloud growth rate is determined we refer to \citetalias{fielding22} and references therein, while here we briefly summarize the basic concepts behind it, all based on results from recent numerical hydrodynamical investigations. As already mentioned above, the main idea is that at the interface (or mixing) layer between the cloud and the hot gas, the material can cool efficiently in a time $\tau_{\rm{cool}}$, equal to the cooling time of the mixed gas, which can be assumed \citep[see][]{gronke18} to have an intermediate temperature $T_{\rm{mix}}=\sqrt{TT_{\rm{cl}}}$ and metallicity (which in our case is always solar given that both phases have the same metallicity value). The growth rate is then given by \cite[see, e.g.,][]{fielding20,gronke20}
\begin{ceqn}
    \begin{equation}
        \dot{m}_{\rm{cl,growth}}=\rho A_{\rm{cool}}v_{\rm{turb}}\xi^{\alpha}\ .
    \end{equation}
\end{ceqn}
Here, $v_{\rm{turb}}$ is the turbulent velocity in the mixing layer, which we approximated as $0.1(v-v_{\rm{cl}})$ following \citetalias{fielding22} (the factor 0.1 comes from the simulations of \citealt{fielding20} and \citealt{tan21}); $A_{\rm{cool}}$ is the cooling area of the cloud; and $\xi=r_{\rm{cl}}/(v_{\rm{turb}}\tau_{\rm{cool}})$ is the parameter that determines whether the system is in a slow-cooling ($\xi<1$, for which $\alpha=1/2$) or a rapid-cooling regime ($\xi>1$, for which $\alpha=1/4$). The cooling area does not coincide with the area of the spherical cloud, since the cloud tends to be elongated due to the interactions with the wind and to form a wake. Assuming that the evolved shape of the cloud is approximated by a cylinder with radius $r_{\rm{cl}}$ and height $(v-v_{\rm{cl}})t_{\rm{cc}}$, where $t_{\rm{cc}}$ is the cloud crushing time, the cooling area is taken to be the lateral surface area of this cylinder, scaled by a correction factor, $f_{\rm{mix}}$ (corresponding to the parameter $f_{\rm{cool}}$ in \citetalias{fielding22}), which represents the efficiency of mixing layer formation (see also Section~\ref{disc:MassExchange}). By using the definition of $t_{\rm{cc}}=\chi^{1/2}r_{\rm{cl}}/(v-v_{\rm{cl}})$, where  $\chi=\rho_{\rm{cl}}/\rho$ is the density contrast between the cold cloud and the hot medium, \citetalias{fielding22} defined the cooling area as
\begin{ceqn}
    \begin{equation}\label{eq:acool}
        A_{\rm{cool}}=f_{\rm{mix}}4\pi r^2_{\rm{cl}}\chi^{1/2}\ ,
    \end{equation}
\end{ceqn}
where $f_{\rm{mix}}$ is left free to vary as a free parameter of our model. Using the definitions outlined above, the final equation describing the cloud mass growth is given by
\begin{ceqn}
\begin{equation}\label{eq:massgrowth}
\dot{m}_{\rm{cl,growth}} = 3 f_{\rm{mix}} \left( \frac{m_{\text{cl}} v_{\rm{turb}}}{\chi^{1/2} r_{\rm{cl}}} \right) \xi^\alpha.
\end{equation}
\end{ceqn}

Finally, to express the overall mass evolution of the cloud, one needs to also take into account its mass losses. The same hydrodynamical instabilities (like Kelvin-Helmoltz instability) that lead to the creation of the mixing layers that might rapidly cool and increase the cloud mass and momentum, may also be responsible for shredding the cloud and making part (or all) of it evaporate into the wind. Following again \citetalias{fielding22}, the mass loss rate can be expressed as
\begin{ceqn}
\begin{equation}\label{eq:massloss_in}
\dot{m}_{\rm{cl,loss}} = \rho_{\rm{cl}}f_{\rm{mix}}4\pi r^2_{\rm{cl}}v_{\rm{turb,cl}}\  ,
\end{equation}
\end{ceqn}
where $f_{\rm{mix}}4\pi r^2_{\rm{cl}}$ is the area over which the cloud is mixed and $v_{\rm{turb,cl}}$ is the turbulent velocity responsible for mixing the cold cloud with the hot wind. Assuming that the turbulent energy density is constant between the cold and hot phases \citep{fielding20}, $\rho_{\rm{cl}}v^2_{\rm{turb,cl}}=\rho v^2_{\rm{\rm{turb}}}$, one can rewrite equation~\eqref{eq:massloss_in} as
\begin{ceqn}
\begin{equation}\label{eq:massloss}
\dot{m}_{\text{cl,loss}} = 3 f_{\rm{mix}} \left( \frac{m_{\text{cl}} v_{\rm{turb}}}{\chi^{1/2} r_{\text{cl}}} \right)\ 
\end{equation}
\end{ceqn}
so that the total cloud mass evolution is given by
\begin{ceqn}
\begin{equation}\label{eq:masstot}
\dot{m}_{\text{cl}} = 3 f_{\rm{mix}} \left( \frac{m_{\text{cl}} v_{\rm{turb}}}{\chi^{1/2} r_{\text{cl}}} \right)(\xi^{\alpha} -1)\ .
\end{equation}
\end{ceqn}

In the right-hand panels of Figure~\ref{fig:meth_orb} we show the effects of the forces described above on the cloud orbits, assuming the same three models analyzed in Section~\ref{sec:modelHot}. For these examples, we assumed a cloud with a mass of $5\times10^2\ M_{\odot}$, starting from  a height $z=50$ pc and a cylindrical radius $R=r_{\rm{inj}}=100$ pc with an initial vertical velocity of $100$ \kms\ and an initial tangential velocity equal to the circular velocity at the starting $R$. For models A and B, we adopted $f_{\rm{mix}}=0.1$ and for model C $f_{\rm{mix}}=0.3$. One can see how in all cases assuming the presence of the entraining hot wind tends to accelerate and push the clouds to much larger distances (solid curves) compared to a simple ballistic model (dotted curves). Moreover, the increase of momentum due to the mixing (equation~\ref{eq:acc_growth}) seems to be the dominant accelerating process and necessary to bring the clouds out to distances of at least $1$ kpc from the disk (dash-dotted curves), while considering only the drag force (equation~\ref{eq:drag}) would bring the clouds out to at most a few hundreds of parsecs (dashed curves). One can also appreciate the difference between the three models: in model C (bottom panel), the high hot gas densities and the higher $f_{\rm{mix}}$ lead the clouds to gain higher velocities and to reach slightly larger distances with respect to the other two models; in models A and B the clouds have instead similar final orbits, meaning that a denser, slower wind has an overall impact on the cloud evolution comparable to that of a faster, less dense one. In all the panels on the right-hand side of Figure~\ref{fig:meth_orb}, we also mark the position of the modeled cloud along its orbit at the beginning (before it is launched), at $5$ Myr and at $10$ Myr.

\begin{figure}
    \centering
    \includegraphics[clip, trim={0cm 0cm 0cm 0cm}, width=0.9\linewidth]{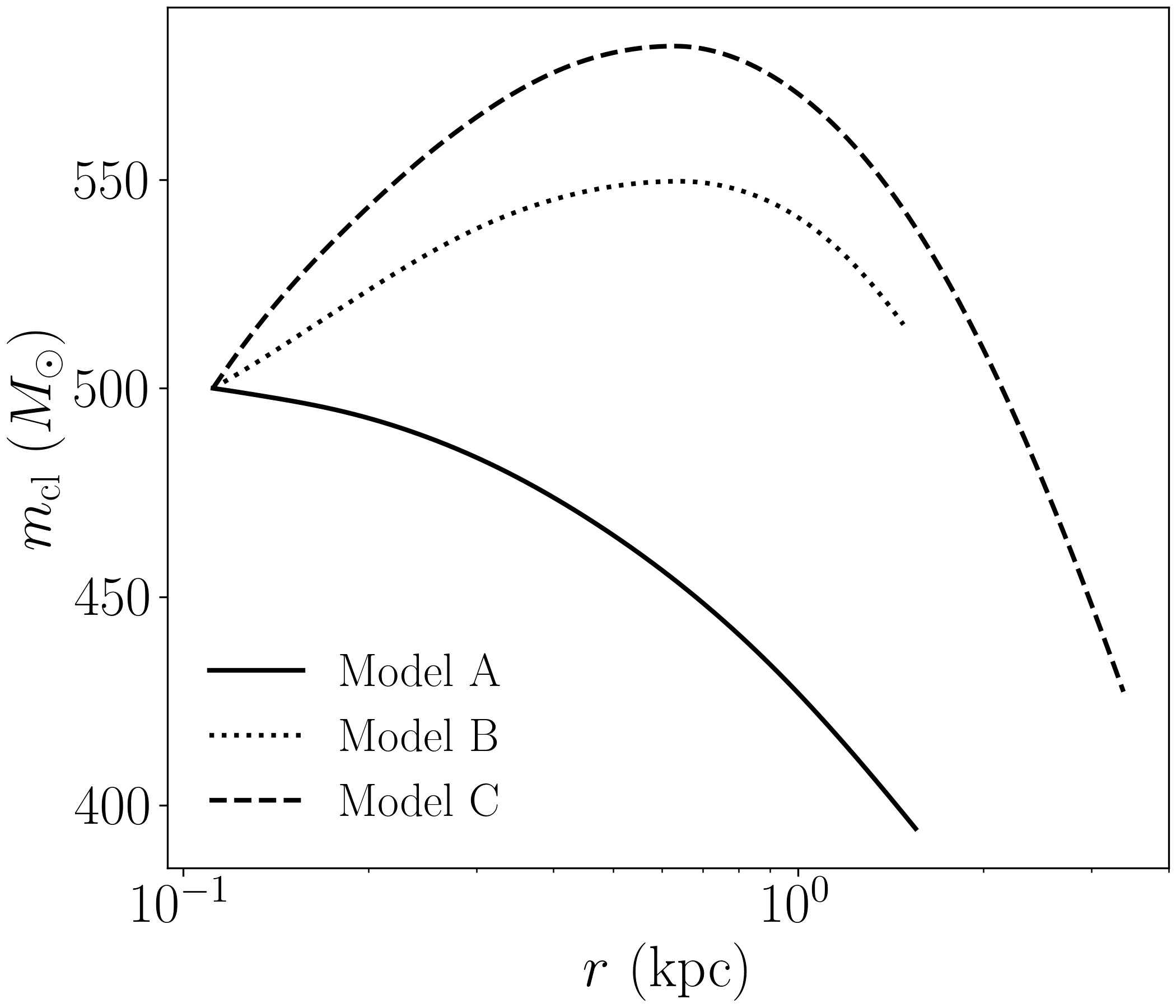}
    \caption{Cloud mass evolution as a function of the distance from the center, for models A, B, and C (see Figure~\ref{fig:meth_orb}). In all cases, the cloud starts with a mass of $5\times10^2\ M_{\odot}$, but in model A it continuously loses its mass and slowly evaporates into the background medium due to hydrodynamical interactions with the hot wind, while in the other two models it initially increases its mass due to the cooling of the mixing layers.}
    \label{fig:meth_mass}·
\end{figure}

The cloud behavior can be better understood by looking at the evolution of the mass of the cloud in the different models, in Figure~\ref{fig:meth_mass}. In model A, even though the cloud is still accelerated as a consequence of momentum gain from the hot wind, it loses more mass to the hot medium than it gains from it at all radii, resulting in a net mass loss ($\xi<1$ everywhere). In model B, the mass growth rate is, in the first part of the orbit, higher than the mass loss rate ($\xi>1$), meaning that the cloud acquires mass from the hot wind. However, the latter is slower than the wind in model A, explaining the similar cloud dynamics between the two models. Finally, in model C the cold mass grows more efficiently than in the previous two models and the cloud acquires gas with velocities close to $1000$ \kms, leading the cloud to a much higher acceleration.

We finally note that, throughout this work, we only considered the effects of the hot wind on the cloud orbits, but for simplicity we did not implement how the exchange of mass and momentum with the clouds affects in return the properties of the hot gas, which are instead fixed once the loading factors and the injection radius are chosen. This is an approximation of the original model of \citetalias{fielding22} that we discuss more in detail in Section~\ref{disc:MassExchange}. 

\subsubsection{Population of cold clouds}\label{sec:population}
In our modeling, we assumed a constant flow of cold clouds that are ejected from the central region of the disk and whose motion outside of the MW disk is driven by the forces described in Section~\ref{sec:modelCold}. To fully define an orbit, one needs to assume a starting position from which the cloud is ejected, the three components of the initial velocity, and a cloud mass. Given that our model is axisymmetric, we only need to choose an initial galactocentric radius, $R_{\rm{in}}$, and height, $z_{\rm{in}}$. We uniformly extracted an initial $R_{\rm{in}}$ between $20\ \rm{pc}$ and the injection radius, $r_{\rm{inj}}$ (see Section~\ref{sec:modelHot}), and $z_{\rm{in}}$ between $\sqrt{r^2_{\rm{inj}}-R^2_{\rm{in}}}$ and $r_{\rm{inj}}$, so that the initial cloud population is effectively distributed on a thin shell outside $r_{\rm{inj}}$ and the clouds are not located inside the sonic region, where the hot wind is still strongly accelerating. We set the initial tangential component of the velocity, $v_{t,\rm{in}}$, equal to the circular velocity at $R_{\rm{in}}$, which we calculated using {\sc galpy} for the adopted MW potential. We then assumed that the initial kick velocity, $v_{\rm{kick}}$, another free parameter of our model, is aligned with the vertical direction, so that $v_{z,\rm{in}}=v_{\rm{kick}}$ and $v_{R,\rm{in}}=0$. We note that one could also assume a random direction for the kick velocity, but we verified a posteriori that this does not have any effect on the results presented in this work. The spherically symmetric hot wind tends indeed to strongly affect the cloud motion and the initial direction of $v_{\rm{kick}}$ has a minimal effect on the final cloud orbit. The kick velocity represents the initial velocity given to the cloud due to its interaction with the hot wind within the injection region, and it should be understood as the result of the cloud being initially swept up by the superbubble created by the SN explosions (which we do not capture in the model). Finally, the initial mass of the cloud is randomly extracted from the observed \hi\ mass distribution, in order to have cloud masses consistent with the data. In particular, we truncated the observed distribution between $30$ and $3000\ M_{\odot}$, to ensure that in the final mass distribution (given that the cloud mass can vary with time, see equation~\ref{eq:masstot}) there would not be clouds with masses lower or higher than those observed.

For a single model realization, we created 20 different orbits, and we followed the cloud motion for $10\ \rm{Myr}$, with an additional stopping criterion in case the cloud gets to a distance $r>2.5$ kpc, given that these distances are not covered by the observational data (Section~\ref{observations}). The orbit integration time was chosen to match the maximum cloud lifetime estimated by \cite{lockman20} with their accelerated outflow model. Indeed, given that we expected our physically motivated models to produce clouds with intrinsic velocities comparable to those predicted by the previous kinematic models (in order to be consistent with the observed cloud kinematics), we also expected comparable cloud lifetimes. We finally populated these orbits with $N_{\rm{cl}}$ clouds\footnote{The total number of ejected clouds $N_{\rm{cl}}$ is relatively arbitrary, given that the likelihood that we used for the Bayesian analysis (see Section~\ref{sec:likelihood} and equation~\ref{eq:likelihood}) is based on the comparison of the shapes of the total cloud distributions of models and data and does not depend on the exact number of clouds. For the Bayesian analysis, we chose $N_{\rm{cl}}=2000$, which ensures that the orbits are sufficiently populated and that the comparison with the data is statistically meaningful. For the plots shown below in Figures~\ref{fig:comparison} and \ref{fig:comparison_new}, we assumed $N_{\rm{cl}}\approx 500$, so that the final number of clouds, after taking into account the observational biases, is $\approx200$, consistent with the number of observed \hi\ clouds.} by placing them at different positions in the orbits assuming that they are evenly separated in time \citep[see][]{afruni21}. Each cloud in the final distribution is uniquely described by $(R_i,z_i,\phi_i,v_{R,i},v_{z,i},v_{t,i},m_{i})$, where the azimuthal angle $\phi_i$ was drawn randomly between $0$ and $2\pi$. We performed this procedure above (positive $z_i$ and $v_{z,i}$) and below (negative $z_i$ and $v_{z,i}$) the plane of the disk.

{ 
\renewcommand{\arraystretch}{1.3}
 \begin{table}
\begin{center}
 \caption[]{Main parameters of our modeling framework.}\label{tab:parameters}
\begin{tabular}{*{3}{c}}
\hline  
\hline
 Parameter & 
 Value & Units\\
\hline 
$SFR$ & 0.1& $M_{\odot}$ yr$^{-1}$ \\
$v_{\rm{c}}$ & 150 & \kms \\
$Z_{\rm{hot}}$ & 1& $Z_{\odot}$ \\
$Z_{\rm{cl}}$ & 1& $Z_{\odot}$ \\
$T_{\rm{cl}}$ & 5000& K \\
\hline
$r_{\rm{inj}}$&$|0.05,0.3|$& kpc\\
$\log \eta_{\rm{E}}$&$|-0.8,0|$& -\\
$\log \eta_{\rm{M}}$&$|-2,0.2|$& -\\
$v_{\rm{kick,100}}$&$|0.3,1.3|$&$100$ \kms\\
$\log f_{\rm{mix}}$&$|-1.5,0|$&-\\

\hline
\end{tabular}
\end{center}
Notes: The parameters $SFR$, $v_{\rm{c}}$, $Z_{\rm{hot}}$, $Z_{\rm{cl}}$, and $T_{\rm{cl}}$ are fixed. The rest of the parameters are free to vary within the flat priors whose limits are shown in the middle column, except for $\log \eta_{\rm{E}}$, for which we adopt a Gaussian prior centered in $-0.4$ and with a width of $0.2$, truncated within the reported boundaries.
   \end{table}
   }

\subsection{Likelihood and priors}
\label{sec:likelihood}
Once the cloud population was created, in order to compare with the observations we converted the cloud positions and velocities, initially in the reference system of the galaxy, to galactic longitudes and latitudes $(l,b)$ and to line-of-sight velocities $v_{\rm{los}}$ (equivalent to $v_{\rm{LSR}}$), using the in-built functions of {\sc galpy}\footnote{In particular, we used \texttt{galpy.util.coords.galcencyl$\_$to$\_$vxvyvz} and \texttt{galpy.util.coords.vxvyvz$\_$to$\_$vrpmllpmbb} for the velocities and \texttt{galpy.util.coords.galcencyl$\_$to$\_$XYZ} and \texttt{galpy.util.XYZ$\_$to$\_$lbd} for the positions.}. We then took into account of the observational biases, by removing clouds that would fall outside of the GBT and the ATCA detectable positions and/or velocities and clouds that would overlap with the \hi\ emission from the MW disk (see Figure~\ref{fig:data} and Section~\ref{observations}).

We then defined our likelihood by directly comparing the final model and observed distributions of galactic latitudes, longitudes, and line-of-sight velocities, through
\begin{ceqn}
\begin{equation}\label{eq:likelihood}
\ln{p_{\rm{tot}}}  = \ln{p_{l}} + \ln{p_{b}} + \ln{p_{v_{\rm{los}}}}\ ,
\end{equation}
\end{ceqn}
where $p_x$ are the probability values of a Kolmogorov-Smirnov (KS) test performed between the observed and predicted cloud properties, i.e.,\ longitude, latitude and line-of sight velocity, respectively. We used the likelihood expressed by equation~\eqref{eq:likelihood} to perform a Bayesian analysis across the 5-dimensional parameter space. As we have seen above, a model is indeed entirely defined by five free parameters: $\eta_{\rm{E}}$ and $\eta_{\rm{M}}$ define how much energy and mass is injected into the hot phase of the wind; $r_{\rm{inj}}$ defines both the radius within which the injection of mass and energy takes place and the initial location of the cold clouds; $v_{\rm{kick}}$ determines the initial ejection velocity of the cold clouds; and $f_{\rm{mix}}$ determines the strength of the mixing between the different phases of the outflow (see Table~\ref{tab:parameters}). We adopted a flat prior on the injection radius $0.05<r_{\rm{inj}}/\rm{kpc}<0.3$, in order to consider radii that are roughly consistent with the extent of the CMZ \citep[see][]{henshaw23}; a truncated Gaussian prior in $\log\eta_{\rm{E}}$ between $-0.8$ and $0$, centered in $-0.4$;\footnote{We adopted a Gaussian prior since we checked that a flat prior would lead to a large and unconstrained posterior distribution in $\log\eta_{\rm{E}}$, but this choice does not significantly affect the posterior distributions of the other free parameters. The reason for the specific choice of this value is discussed in Section~\ref{sec:Propbestfit}.} a flat prior in $\log{\eta_{\rm{M}}}$ between $-2$ and $0.2$; a flat prior in the kick velocity $0.3<v_{\rm{kick,100}}<1.3$ (where $v_{\rm{kick}}$ is expressed in units of $100$ \kms); and a flat prior in the normalization of the mass exchange, $-1.5<\log{f_{\rm{mix}}}<0$.

\section{Results}\label{ModelResults}

\begin{figure}
    \centering
    \includegraphics[clip, trim={0.26cm 0.23cm 0.2cm 0.2cm}, width=\columnwidth]{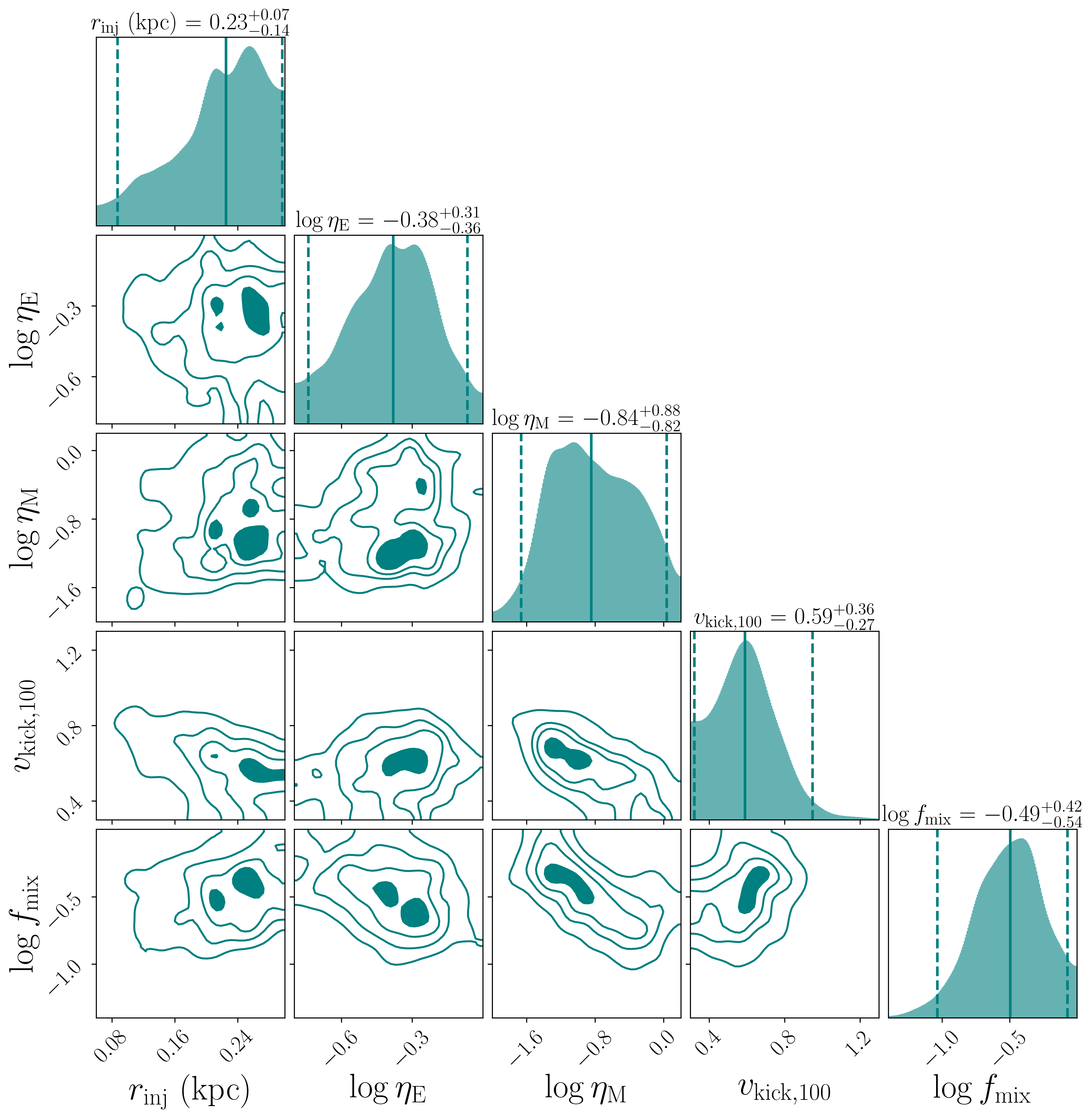}
    \caption{Corner plot showing the posterior distributions of the five free parameters of our models. The vertical lines mark the positions of the 2.5, 50 (solid line) and 97.5 percentiles of the one-dimensional posterior distributions.}
    \label{fig:cornerMultiph}
\end{figure} 

\begin{figure}
    \centering
    \includegraphics[clip, trim={0cm 0cm 0cm 0cm}, width=\linewidth]{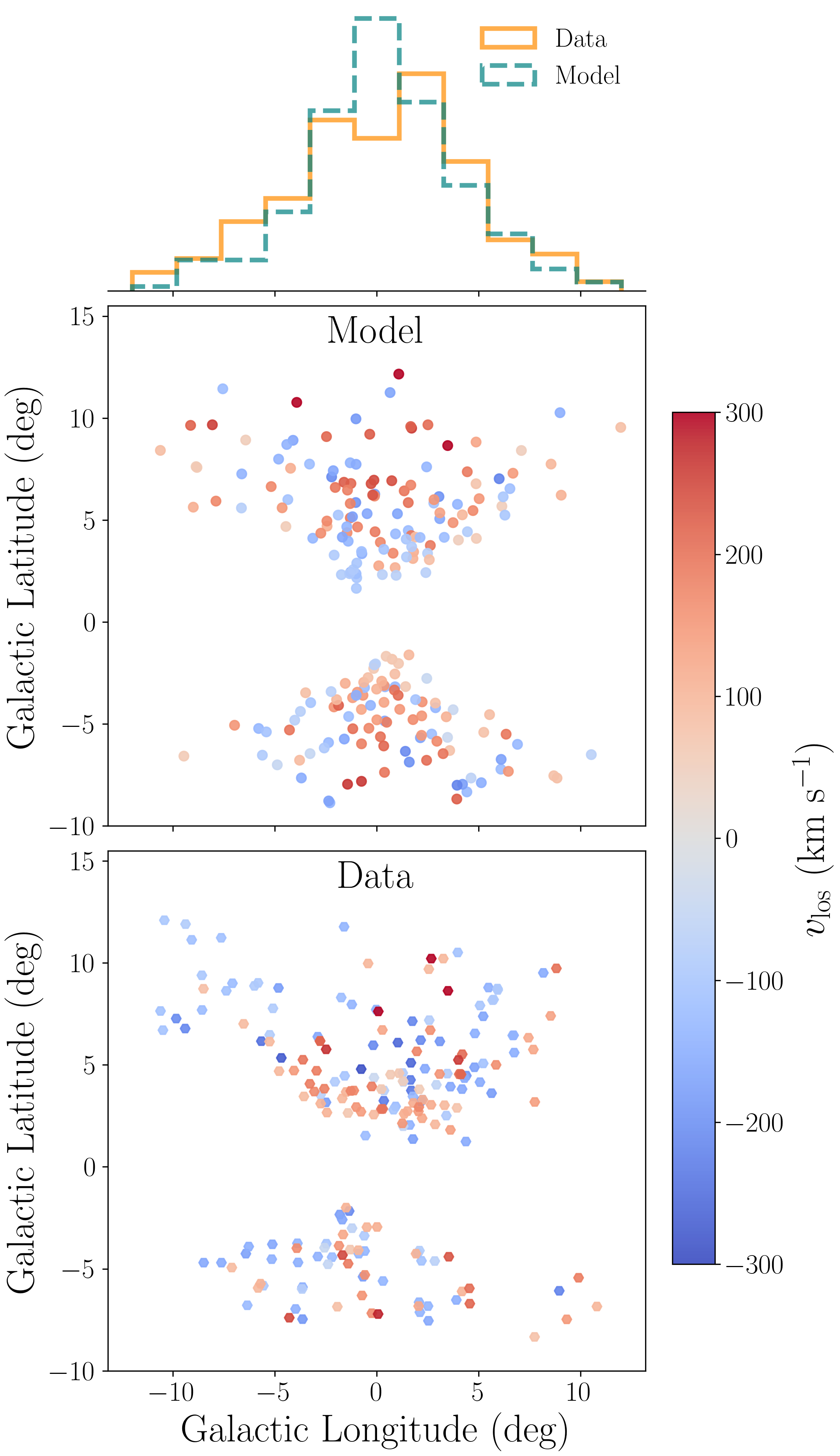}
    \caption{Comparison of the outputs of the best-fit models with the observational data. The cloud kinematics is shown as a function of Galactic longitude and latitude, with models on the top and data on the bottom. The one-dimensional distributions of the observed (orange, solid line) and the model (teal, dashed line) longitudes are reported on top.}
    \label{fig:comparison}
\end{figure} 

\begin{figure}
    \centering
    \includegraphics[clip, trim={0.3cm 0cm 3cm 2.7cm}, width=\linewidth]{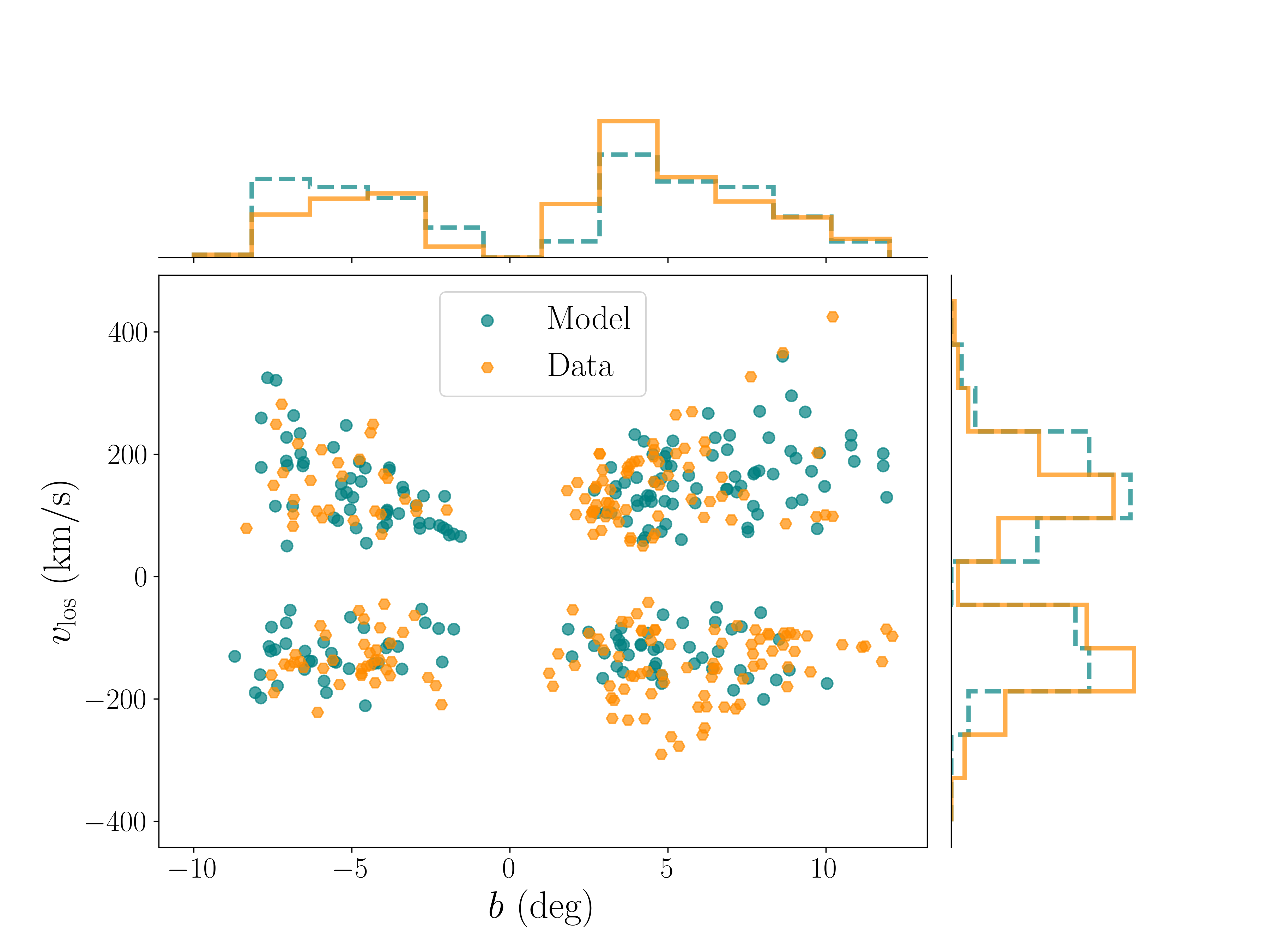}
    \caption{Comparison of the outputs of the best-fit models with the observational data. The model (teal) and data (orange) line-of-sight velocities are shown as a function of the Galactic latitude. The marginal one-dimensional distributions are plotted along the two axes for both the data and the model.}
    \label{fig:comparison_new}
\end{figure} 

\subsection{Bayesian analysis and comparison with the data}
To perform the Bayesian fit, we adopted the nested sampling method \citep{skilling04,skilling06}, using the {\sc dynesty} python package \citep[][]{speagle20,koposov22}. The posterior distributions of the five free parameters (see Section~\ref{sec:likelihood}) are shown in Figure~\ref{fig:cornerMultiph}. While we discuss below the meaning of these best-fit values (Section~\ref{sec:Propbestfit}), it is immediately evident how there is a well-defined region of the parameter space that better reproduces the data outlined in Section~\ref{observations}.  
The median values of the five posterior distributions are: $r_{\rm{inj}}=0.23$ kpc, $\eta_{\rm{E}}=0.42$, $\eta_{\rm{M}}=0.14$, $v_{\rm{kick}}=59$ \kms, $f_{\rm{mix}}=0.32$. The most prominent degeneracies are present between the hot gas mass loading factor and the kick velocity and the mixing normalization: less massive hot winds require cold clouds with larger initial $v_{\rm{kick}}$ and/or higher $f_{\rm{mix}}$ in order to produce outputs more consistent with the observations.

We first investigate the goodness of the best-fit models in reproducing the observed kinematics of the \hi\ clouds. This can be appreciated in Figures~\ref{fig:comparison} and \ref{fig:comparison_new}, where we compare the data directly with the outputs of a model with the five free parameters fixed to the median values of the posterior distributions. In Figure~\ref{fig:comparison}, one can see the cloud line-of-sight velocity maps in galactic longitude and latitude for the best-fit model (top) and the data (bottom). Even though the distributions present some expected differences, we note that the general kinematic pattern is very well reproduced by our model, both in terms of velocity and of cloud location across the sky. The two distributions above the top panel additionally show the normalized distributions of longitudes for model (teal, dashed) and data (orange), whose comparison is used in the first term of the likelihood in equation~\eqref{eq:likelihood}.

The consistency between model outputs and data is even more clear by the comparison in Figure~\ref{fig:comparison_new} between the latitudes of the clouds as a function of their line-of-sight velocities (model in teal and data in orange). The modeled clouds show a sign of acceleration with increasing distance from the Galactic plane, very closely following the pattern detected in the \hi\ data (see right panel of Figure~\ref{fig:data} and \citealt{lockman20}). This is confirmed also by the comparison of the 1-d distributions of latitudes on the $x$-axis (corresponding to the second term of the likelihood, equation~\ref{eq:likelihood}) and of line-of-sight velocities on the $y$-axis (third term in equation~\ref{eq:likelihood}).

We can therefore conclude that we found a class of physically motivated models that lead to a cloud population in good agreement with the observations of Section~\ref{observations}, in terms of velocities, cloud locations and cloud masses (by construction, given that in the models we extract cloud masses from the observed mass distribution, see Section~\ref{sec:population}). In the next section, we explore more in-depth the physical properties of the multiphase wind predicted by our best-fit models.

\subsection{Properties of the best-fit models}\label{sec:Propbestfit}
\begin{figure}
    \centering
    \includegraphics[clip, trim={0cm 0cm 0cm 0cm}, width=\linewidth]{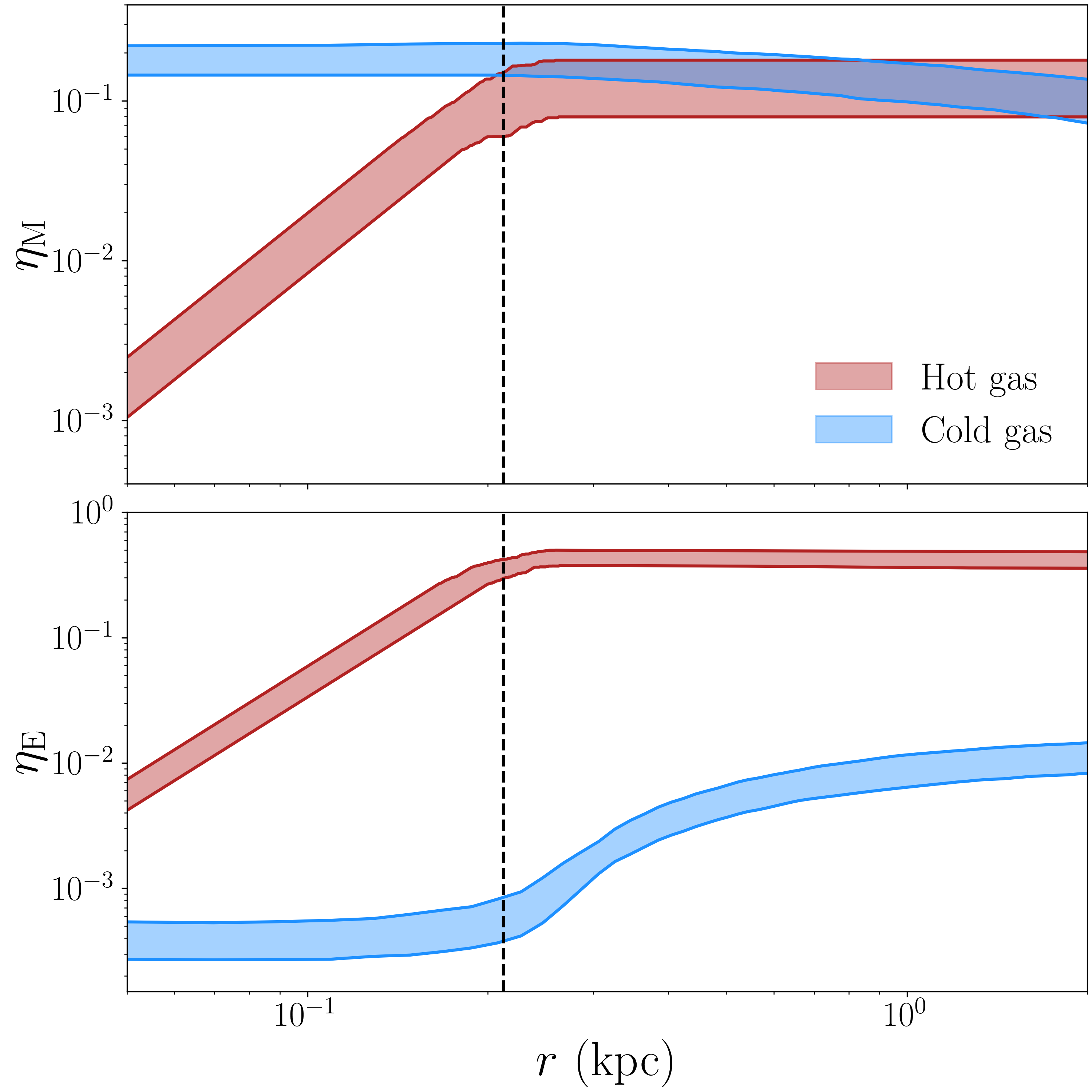}
    \caption{Mass (top) and energy (bottom) loading factors as a function of the galactocentric radius, $r$, for a class of best-fit models (see Section~\ref{sec:Propbestfit} for more details), for the hot gas (red) and the cold clouds (blue). The vertical dashed line marks the location of the median injection radius.}
    \label{fig:Lfactors}
\end{figure}

We first focus on the hot gas phase and specifically on the mass and energy loading factors of the hot wind. These are shown in red respectively in the top and bottom panels of Figure~\ref{fig:Lfactors} as a function of the spherical galactocentric radius, $r$, and are given by

\begin{ceqn}
\begin{equation}\label{eq:etaM}
\eta_{\rm{M}}(r)=\frac{\dot{M}_{\rm{out}}(r)}{SFR}= \frac{4\pi r^2\rho(r) v(r)}{SFR}\ ,
\end{equation}
\end{ceqn}

\begin{ceqn}
\begin{align}\label{eq:etaE}
\eta_{\rm{E}}(r)=\frac{\dot{E}(r)}{E_{\rm{SN}} SFR/(100\ M_{\odot})}= \nonumber \\ \nonumber\\
~ \frac{4\pi r^2\rho(r) v(r) \left[ \frac{1}{2}v^2(r) +\frac{3}{2}c^2_{\rm{s}}(r)\right]}{E_{\rm{SN}} SFR/(100\ M_{\odot})}\ ,
\end{align}
\end{ceqn}
where the two terms in equation~\eqref{eq:etaE} take into account respectively the kinetic and thermal energy outflow rates.

The width of the bands in Figure~\ref{fig:Lfactors} represents the $1-\sigma$ interval of $200$ models with parameters extracted from the posterior distributions shown in Figure~\ref{fig:cornerMultiph}. Both loading factors increase up to $\sim150-200$ pc, corresponding to the injection radius (the median value is shown by the black dashed vertical line), within which we inject by construction all the energy and mass coming from the supernovae. At $r_{\rm{inj}}$ both factors therefore reach a maximum, with $\eta_{\rm{E}}$ being roughly equal to 0.4 and $\eta_{\rm{M}}\sim0.1$. Both loading factors remain then constant at larger radii: while this is expected by construction for the mass loading, the roughly constant $\eta_{\rm{E}}$ indicates that the radiative losses of the hot wind are negligible for this class of models. It is important to note that the values of the two loading factors are consistent with the findings of recent high-resolution hydrodynamical simulations of an ISM patch, like TIGRESS \citep[e.g.,][]{kim18,kim20,kim20b}. In terms of star formation rate surface density, the Galactic center is most similar to the TIGRESS R2 model, with $\Sigma_{SFR}\approx1\ M_{\odot}\ \rm{kpc}^{-2}\ \rm{yr}^{-1}$, for which they found an energy loading factor $\eta_{E}\approx0.6$ in the central regions and then decreasing with the distance from the disk. This justifies our choice of introducing a Gaussian prior at a value $\eta_{\rm{E}}\sim0.4$ in our Bayesian analysis. As for the hot gas mass loading factor, it has been found to be relatively independent of the central star formation rate and to be of the order of $10\%$ \citep[see][]{kim20,schneider20}: we stress that our recovered value of $\eta_{\rm{M}}$, which is only set by the comparison with the observations, is interestingly very similar to this value and in very good agreement with these theoretical findings.

We discuss in Section~\ref{disc:xraydata} the rest of the properties of the hot phase, but given the value of the loading factors, we can already conclude that the hot wind in our best-fit models is physically justified and consistent with expectations from state-of-the-art theoretical models.

\begin{figure*}
    \centering
    \includegraphics[clip, trim={0cm 0cm 0cm 0cm}, width=\linewidth]{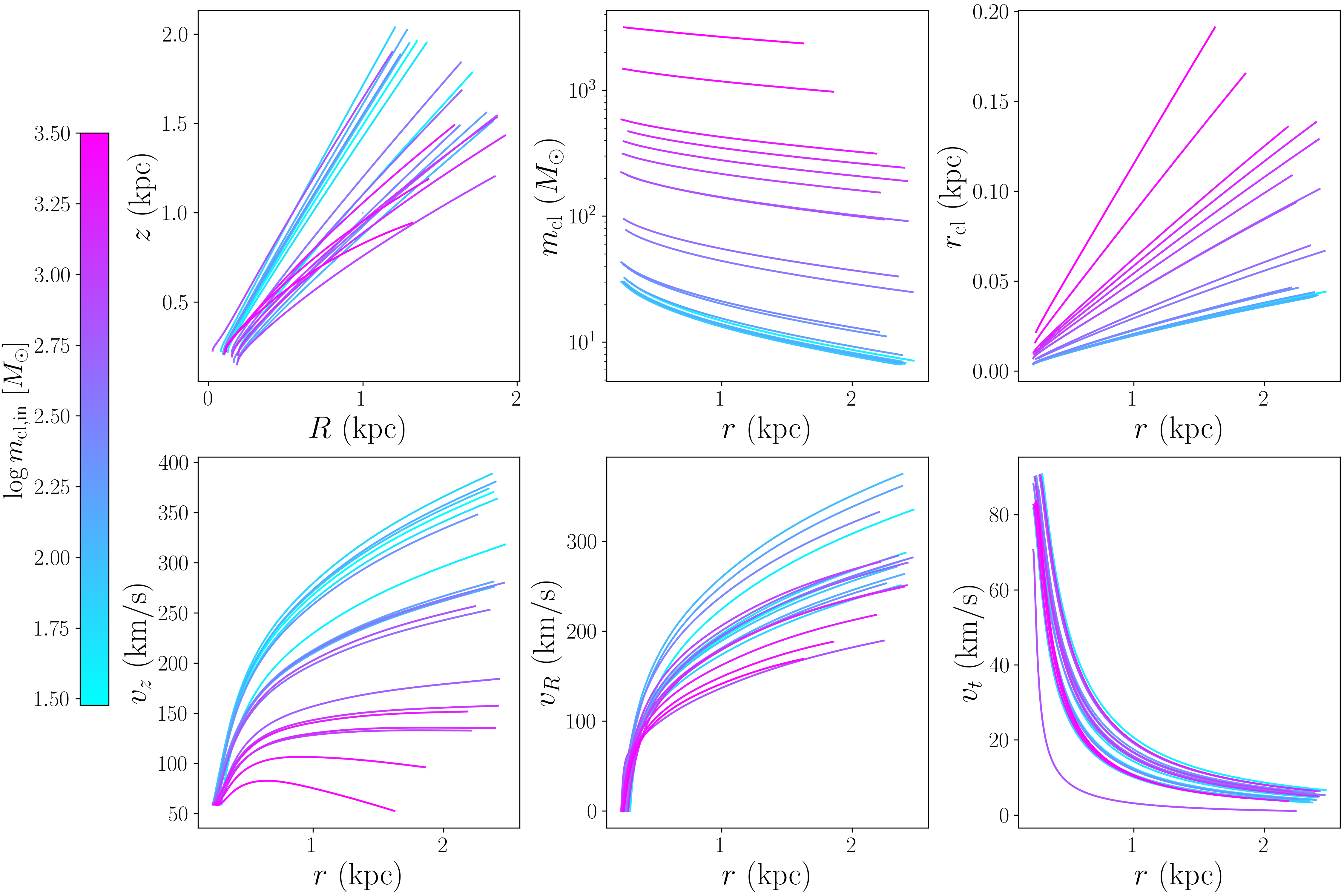}
    \caption{Intrinsic properties of the \hi\ clouds predicted by our best-fit models. Top, from left to right: Cloud orbits, cloud masses as a function of $r$, and cloud radii as a function of $r$. Bottom: Three components of the cloud velocities. The different curves are colored based on the initial cloud mass, from lower masses in cyan to higher masses in magenta.}
    \label{fig:CloudProperties}
\end{figure*}

In Figure~\ref{fig:CloudProperties} we report instead the intrinsic properties of the \hi\ clouds predicted by our best-fit models. The different curves show the cloud properties along 20 different orbits that were created fixing the values of the five free parameters to the respective median values of the posterior distributions. The curves are colored based on the initial cloud mass, from cyan (lower mass clouds) to magenta (higher mass clouds). The actual orbits are shown in the upper left panel, where one can see that the clouds start from a thick shell close to the center of the MW (at about $r_{\rm{inj}}$) and are then ejected and entrained by the hot wind at distances of $1.5-2$ kpc from the center, consistent with the observations. Clouds with lower masses ($\lesssim10^2\ M_{\odot}$) tend to be more affected by the hot wind and are entrained up to larger distances with respect to more massive clouds ($\gtrsim10^3\ M_{\odot}$), which instead are generally not able to reach heights higher than about $0.8-1$ kpc within 10 Myr. This effect can also be noted by looking at the cloud velocity components (bottom panels): while we can see that all the clouds get accelerated in the radial direction as they increase their distance from the center, the vertical velocities of high-mass clouds never exceed $100\ \rm{km}\ \rm{s}^{-1}$ and actually tend to decelerate from the initial kick velocity. We have seen in Section~\ref{observations} (Figure~\ref{fig:data}) that in the data there is no strong dependence of the \hi\ kinematics with the estimated cloud mass, but that most of the clouds with very high velocity have low masses, which is consistent with this emerging feature of our models. As for the tangential velocities (bottom right panel of Figure~\ref{fig:CloudProperties}), the clouds are ejected by construction with the MW circular velocity at the respective initial galactocentric radius, $R_{\rm{in}}$, going from about $70$ to $90\ \rm{km}\ \rm{s}^{-1}$. Because of the gain of momentum from the hot wind (which is by construction non-rotating) the tangential velocities tend then to quickly drop and have values of at most $20\ \rm{km}\ \rm{s}^{-1}$ at $r>1$ kpc. This feature is also consistent with the observed kinematic pattern of the \hi\ cloud population, which does not show any sign of global rotation (see Section~\ref{observations} and \citealt{diteodoro18}). 

The central top panel of Figure~\ref{fig:CloudProperties} shows the evolution of the cloud masses. One can see how, with this choice of parameters, all clouds tend to lose part of their mass into the hot wind, meaning that the mass loss term dominates in equation~\eqref{eq:masstot} and that $\xi<1$ at all radii. In particular, smaller clouds tend to be disrupted more rapidly and at $2$ kpc have lost more than $70\%$ of their initial mass. Therefore, although the clouds do accrete some high-velocity material, which contributes to their acceleration, they experience a net mass loss to the wind. This is an important result of our work: even though our framework allows the clouds to lose or gain mass, in order to reproduce the observed kinematics the clouds have to overall evaporate (i.e., lose mass due to stripping of material from the interactions with the background medium) into the hot wind as they move away from the disk. We therefore speculate that these clouds would likely not be able to reach distances from the MW disk larger than a few kiloparsecs. 

We calculated the average mass lost by the clouds at a radius $r$, $f_{\rm{loss}}(r)= \langle m_{\rm{cl} }(r)\rangle/\langle m_{\rm{cl,in}}\rangle$, in order to calculate the average cold gas mass loading factor as a function of radius, assuming that the $N_{\rm{cl}}$ clouds in our model have been ejected for $10$ Myr at a constant rate:

\begin{ceqn}
\begin{equation}\label{eq:etaMcold}
\eta_{\rm{M,cold}}(r)=\frac{N_{\rm{cl}}\langle m_{\rm{cl,in}}\rangle f_{\rm{loss}}(r) / (10\ \rm{Myr})}{SFR}\ .
\end{equation}
\end{ceqn}
In a similar fashion, the energy loading factor of the cold gas was instead given by

\begin{ceqn}
\begin{equation}\label{eq:etaEcold}
\eta_{\rm{E,cold}}(r)=\frac{\eta_{\rm{M,cold}}(r)SFR\left[ \frac{1}{2}v_{\rm{cold}}^2(r) +\frac{3}{2}c^2_{\rm{s,cold}}(r)\right]}{E_{\rm{SN}} SFR/(100\ M_{\odot})}\ ,
\end{equation}
\end{ceqn}
where $v_{\rm{cold}}=\sqrt{v_R^2 + v_z^2}$, which excludes the tangential motion that is not related to the wind. In order to get realistic loading factors, we chose a value of $N_{\rm{cl}}$ that would give us a final number of clouds, after taking into account of all the observational biases (see Section~\ref{sec:likelihood}) consistent with the observed one (213). The loading factors are shown in blue in Figure~\ref{fig:Lfactors}, with the width of the bands representing the $1-\sigma$ uncertainty calculated in the same way as for the hot gas. One can see that the cold gas mass loading factor is slightly larger than that of the hot phase, especially in the inner regions,\footnote{Outside of the injection radius, given that by construction we only insert clouds at initial distances that are $\geq r_{\rm{inj}}$, so the cold loading factors at lower distances are just an extrapolation.} while it decreases at larger radii due to the cloud mass losses into the hot wind. The energy loading factor is instead almost two order of magnitudes lower than that of the hot wind. This picture is in agreement with high-resolution magnetohydrodynamical simulations of patches of ISM, where the cold outflow usually dominates in mass, with loading factors lower than unity (and decreasing with height) but generally higher than those of the hot phase, while the hot wind dominates the outflow energy budget \citep[e.g.,][]{li20}. However, we note that our calculation of the cold loading factors was based on the total number of ejected clouds over time and might not be directly comparable with the loading factors extracted from simulations \citep[e.g.,][]{kim20, schneider20}, which come directly from the ratio between the mass flux (or outflow rate) and the star formation rate surface density (or star formation rate). Moreover, our value represents only a lower limit, given that it depends on the number of detected \hi\ clouds: as already mentioned, the number $N_{\rm{cl}}$ used in equation~\eqref{eq:etaMcold} was indeed chosen such that after taking into account the observational biases discussed above (see Section~\ref{sec:likelihood}) the number of `detected' clouds would be similar to the observed one (213); if many \hi\ clouds are currently missed by the observations due to biases not taken into account in our analysis, then the number $N_{\rm{cl}}$, and consequently the cold gas mass loading factor, would be higher.

Finally, the top-right panel of Figure~\ref{fig:CloudProperties} shows the radii of the clouds as a function of the distance, $r$, from the Galactic center. Even though the clouds are losing mass due to the interactions with the hot wind, they are increasing in size given that we assumed pressure equilibrium between the two gas phases and that the pressure of the hot gas drastically drops with increasing $r$. Therefore, $r_{\rm{cl}}$ goes from less than $10$ pc close to the injection radius up to more than $100$ pc for the most massive clouds at distances between 1 and 2 kpc. We note that these sizes are slightly larger than those estimated for the observed \hi\ clouds, where the largest cloud radius is approximately $50$ pc \citep[see][]{diteodoro18}. This could be explained by at least some of these clouds being out of pressure equilibrium. The distribution of timescales traveled by the clouds in the final cloud population of our best-fit models goes from 1 to 10 Myr,\footnote{Note that these timescales are not necessarily set by our choice of the maximum integration time. If the fit had, for example, preferred models with much higher velocity clouds, most of the `detected' clouds would have much shorter timescales than what was found here.} with a peak around 5 Myr (consistent with the previous estimate from \citealt{lockman20}), while the cloud crossing time is between 2-15 Myr, meaning that some of the clouds could still not have reached an equilibrium state. In addition to this, the discrepancy between the observed and modeled cloud sizes might also be due to uncertainties in the density profile of the wind, which we discuss more in detail in Sections~\ref{disc:WindAssumptions} and~\ref{disc:MassExchange}.

\begin{figure}
    \centering
    \includegraphics[clip, trim={0.3cm 0cm 0cm 0cm}, width=\columnwidth]{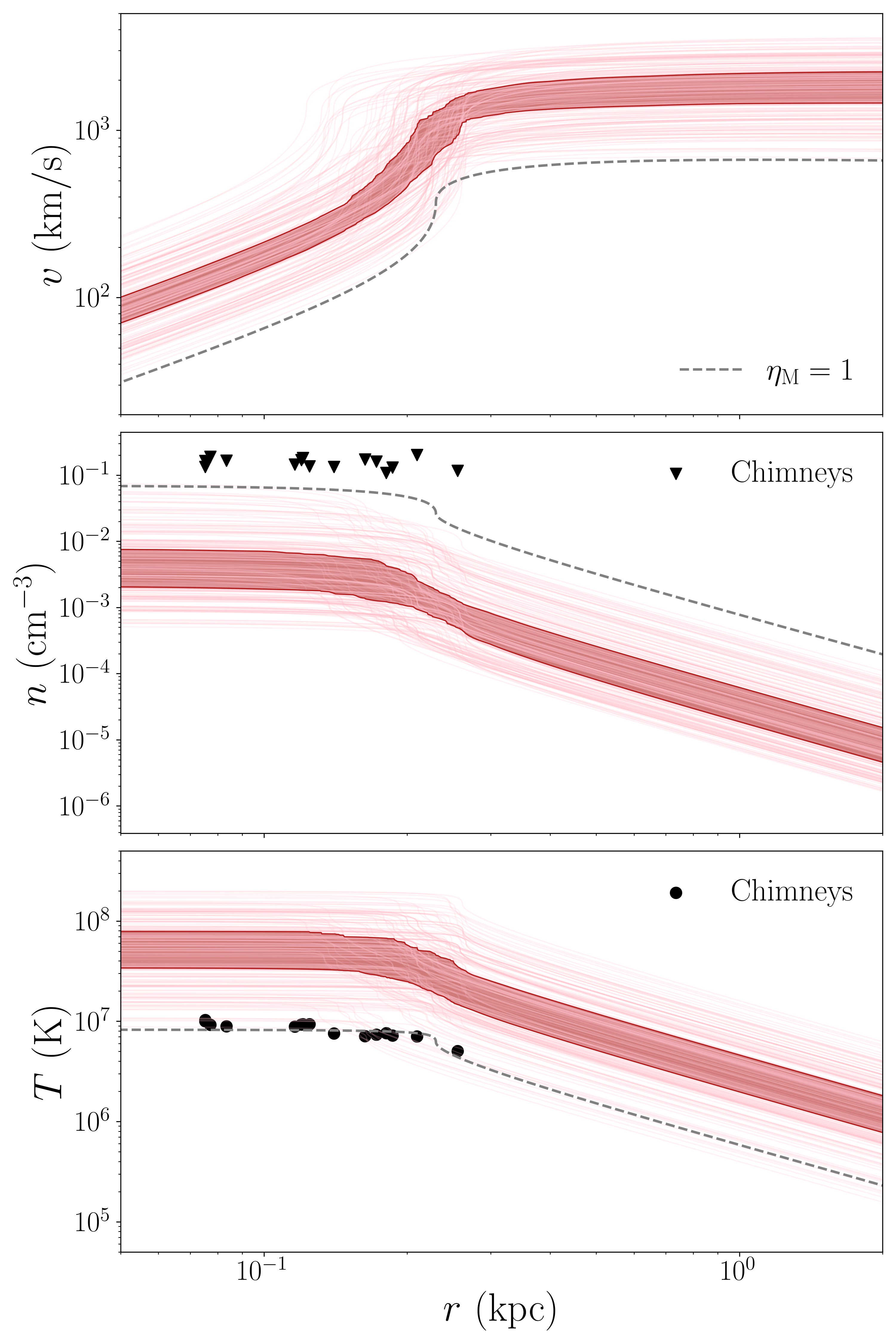}
    \caption{Properties of the hot wind predicted by our best-fit models (in red; see main text for more details), compared to recent X-ray observational constraints from XMM-Newton \citep{ponti19}. From top to bottom, hot gas velocity, density, and temperature. The dashed curves show the predictions for a model where the hot mass loading factor $\eta_{\rm{M}}$ is forced to values close to 1 (see Section~\ref{disc:xraydata}).}
    \label{fig:HotgasProp}
\end{figure} 

\section{Discussion}\label{discussionSection}
We have found in Section~\ref{ModelResults} that there is a class of physically motivated models, consistent with various theoretical expectations, that predict \hi\ cloud properties and kinematics in good agreement with the available constraints from 21-cm data of a population of high-velocity clouds close to the Galactic center. In this section, we explore the comparison of our model predictions with additional observational data of the multiphase MW wind (Section~\ref{disc:comparisonObs}) and we discuss more in detail the assumptions and uncertainties of our modeling (Sections~\ref{disc:WindAssumptions} and \ref{disc:MassExchange}).

\subsection{Comparison with multiwavelength observations} \label{disc:comparisonObs}
\subsubsection{X-ray data}\label{disc:xraydata}

Although the main aim of this work is to infer and interpret the dynamics of the cold \hi\ clouds, the hot phase of the wind is a fundamental ingredient of our physical picture. The motion of the clouds is strongly influenced by the interaction with the hot gas and, without accounting for this phase, it would not be possible to accurately reproduce the observations
(see Appendix~\ref{sec:ballistic} for more details). We have already discussed above (Section~\ref{sec:Propbestfit}) that the hot diffuse wind has mass and energy loading factors that are in agreement with theoretical expectations. Here, we compare its properties with the observational constraints currently available from X-ray data.

In the three panels of Figure~\ref{fig:HotgasProp} we show the intrinsic properties (velocity, density, and temperature) of the best-fit hot wind predicted by our Bayesian analysis, as a function of the radius, $r$. The pink curves show the outputs of 200 models with parameters taken from the posterior distributions shown in Figure~\ref{fig:cornerMultiph} between the 16$^{\rm{th}}$ and 84$^{\rm{th}}$ percentiles, while the red bands show the $1-\sigma$ scatter at each radius. One can see that within the injection radius the wind is still accelerating and has densities of about $5\times10^{-3}\ \rm{cm}^{-3}$ and temperatures between $10^7$ and $10^8$ K. At the injection radius, the wind reaches then a terminal velocity of about $1500$ \kms\ and its density and temperature start to drop to reach values of about $10^{-5}\ \rm{cm}^{-3}$ and $10^6$ K at $r\approx2\ \rm{kpc}$.

While current X-ray observations do not have the necessary spectral resolution to infer hot gas velocities, we can already compare our densities and temperatures with observational estimates of the hot material detected in the vicinity of the galactic center. A few years ago, using XMM-Newton, \cite{ponti19} discovered X-ray structures extended up to hundred-parsec scales from the center, which they named Galactic Center Chimneys \citep[see also][]{nakashima19}. These are almost cylindrical hot plasma structures that extend more than 1 degree in latitude from the Galactic disk; they could be the channel of energy transport from the Galactic center to the Fermi/eRosita bubbles and might be powered by either SN explosions or activity of Sagittarius A$^{\star}$.

Given their properties, these chimneys could potentially be related with the inner part of the hot wind predicted by our model. In the second and third panels of Figure~\ref{fig:HotgasProp}, we report in black the observational estimates of the chimneys' temperature and densities, which span a range in galactocentric radii from about $70$ to $200$ pc \citep[see][]{ponti19}. Our models predict a hot gas phase that is on average hotter and less dense than these observations.\footnote{It should be noted though that the observational densities are only upper limits, given that they are based on assumptions on the three-dimensional geometry of the chimneys, which is still unknown. It is also worth noting that part of the X-ray emission associated with the chimneys could come from foreground sources \citep[e.g.,][]{wang21}.} This might imply that with our models we are constraining a different component, which is still too faint to be detected by current X-ray facilities. 

To further investigate the discrepancy between our models and the data discussed above, we ran an additional Bayesian fit where we fixed the mass loading factor to values close to 1 (imposing an additional Gaussian prior in $\eta_{\rm{M}}$), in order to force the wind to higher densities, more compatible with the data from \cite{ponti19}. The median values of the resulting posterior distributions are: $r_{\rm{inj}}=0.23$ kpc, $\eta_{\rm{E}}=0.54$, $\eta_{\rm{M}}=0.98$, $v_{\rm{kick}}=44$ \kms, $f_{\rm{mix}}=0.22$ (see also Figure~\ref{fig:corneretaM0}). The hot gas outputs of such model are shown by the dashed curves in Figure~\ref{fig:HotgasProp}: the hot gas is slower ($\sim700$ \kms) than in our fiducial models and has inner temperatures of about $10^7$ K, also in better agreement with the X-ray data. The resulting cloud kinematics is still in a relatively good agreement with the observed one, even though it is not as accurate as for our fiducial models (see Appendix~\ref{sec:massiveWind} for more details) . We stress that, to produce a similar wind, we need a large mass loading factor of about 1, which is higher than what is typically predicted by high-resolution simulations \citep[e.g.,][]{kim20, schneider20, rathjen21}. 
$\eta_\mathrm{M}$ enters the model through the mass injection rate, defined as $\dot{M}_\mathrm{inj} = \eta_\mathrm{M} SFR$, where we fix $SFR = 0.1\ M_{\odot}\ \mathrm{yr}^{-1}$, consistent with observational estimates. However, the exact value of the $SFR$ is still debated (see Section~\ref{disc:WindAssumptions}). If a higher $SFR$ were assumed, a lower value of $\eta_\mathrm{M}$ would be required to reproduce hot wind properties consistent with X-ray observations. 
Overall, we conclude that our framework allows, taking into account all the observational and theoretical uncertainties, for a denser and cooler wind that would potentially be more in agreement with the X-ray constraints.

At Galactic latitudes higher than 1 degree, X-ray emission has recently been detected by eRosita \citep{predehl21}. The analysis of the eRosita spectra has shown that the X-ray emission is likely due to the combination of multiple components, some related to Galactic scales, like the eRosita bubbles and the MW hot circumgalactic medium \citep[CGM, see][]{ponti23,locatelli24} and others related to more local sources like the local hot bubble \citep{yeung24}. 
We note that to qualitatively better reproduce the emission measures of the Galactic X-ray components detected by eRosita, models with large mass loading factors (like those described above) are preferred to our fiducial best-fit models, which instead predict a less dense hot wind. However, given the uncertainties in both the observations and our model assumptions (see Sections~\ref{disc:WindAssumptions} and \ref{disc:MassExchange}), we leave a more thorough comparison with the eRosita observations for future work.

\subsubsection{UV data}\label{Uvdata}
As already mentioned in Section~\ref{intro}, the cooler phases of the MW outflow have been detected not only through \hi\ emission, but also extensively through UV absorption lines in the spectra of background quasars (QSOs) and halo stars \citep[e.g.,][]{fox15,bordoloi17,cashman23}. While the \hi\ data are confined to latitudes lower than $10$ degrees \citep[but see][]{bordoloi25}, the sightlines where UV absorption was detected are all located at $10^{\circ}<|b|<50^{\circ}$, meaning that the two surveys trace two different areas of the MW wind, making it more challenging to assess whether the cold and warm phases of the outflow share the same physical origin. 

\cite{bordoloi17} find that the velocities of the warm ($T\sim10^{4-5}$ K) clouds traced by the UV absorption go from about 250 \kms\ at about 2.5 kpc from the disk down to 90 \kms\ at a distance of 6.5 kpc. They model such kinematic signature with models of a decelerating outflow that is launched with initial intrinsic velocities of about $1000$ \kms.  
Our findings, with cold clouds that are accelerated by a hot fast wind and reach radial intrinsic velocities of about $300$ \kms\ at $1-2$ kpc (see Figure~\ref{fig:CloudProperties})\footnote{Similar velocities were also found using simpler kinematic outflow models made to reproduce the \hi\ high-velocity clouds, see \citealt{diteodoro18,lockman20}.}, therefore predict a distinct scenario compared to the outflow necessary to reproduce the warm clouds. This could mean that the two phases (warm and cold gas) might have a different kinematics, with that of the warm medium being potentially more similar to the kinematics of the hot wind. On the other hand, assuming that the clouds predicted by our models would survive up to heights larger than 2 kpc (but see Section~\ref{sec:Propbestfit}) and would experience a deceleration when interacting with the pre-existing material (see Section~\ref{disc:WindAssumptions}), our physical picture could potentially be able to reproduce at the same time both UV and 21-cm data. We plan to investigate this in a future study.

\subsection{Main assumptions of the wind model}\label{disc:WindAssumptions}
A fundamental assumption of this analysis is that the MW wind is powered by SN feedback. As already mentioned in Section~\ref{intro}, it is still debated whether the processes that originate the Fermi Bubbles and potentially also the wind entraining the cold \hi\ clouds, are related to star formation or to the activity of Sgr A$^{*}$ \citep[e.g.,][]{yangRuzkowski22}. The model of \citetalias{fielding22}, while developed for SN-driven winds, can in principle also be applied to an AGN-driven wind, by accordingly changing the injections of mass and energy in order to reproduce the expectations from the central SMBH. It is important to note that the model of \citetalias{fielding22}, originally based on the model of \cite{chevalier85}, describes a steady-state wind, while AGN winds are blast wave-like \citep[][]{sedov46,taylor50} and are usually described by models that follow the time-dependent evolution of a shell of swept-up gas \citep[see, e.g.,][]{king03,giguere12,king15,richings18b,marconcini25}. While we acknowledge that this could be another channel to reproduce the observations of Section~\ref{observations}, developing such models for the MW wind is outside the scope of this work.

Throughout this paper, we have assumed that the wind is powered by the SN explosions within a spherical region at the center of the MW. Through equations~\eqref{eq:etaM}-\eqref{eq:etaEcold}, we relate the mass and energy loading factors of the hot and the cold phase to the star formation rate. In our models, we assume $SFR=0.1\ M_{\odot}/\rm{yr}$: this is consistent with observational estimates of the current star formation rate in the CMZ from both direct counting of young stellar objects and integrated light measurements \citep[e.g.,][]{barnes17} and is expected to be relatively constant for at least the last $5$ Myr and likely for a longer timescale \citep[see][]{henshaw23}. While our assumption is therefore justified, variations of the adopted $SFR$ by a factor of a few and a possible evolution throughout the $10$ Myr analyzed in our work cannot be excluded \citep[e.g.,][]{noguerasLara20,noguerasLara22}. Such differences would most likely lead to slightly different values for the mass and energy loading factors that we inferred in Section~\ref{ModelResults}. In addition to the value of the star formation rate, our model has the simplifying assumption of a spherical injection of mass and energy into the hot phase of the wind, while this region would be more realistically represented by a ring, resembling the structure of the CMZ \citep[e.g.,][]{armillotta19,Sormani20,Tress20}. A more sophisticated axisymmetric model for the hot gas phase \citep[e.g.,][]{sofue19,nguyen21,nguyen22} would naturally predict biconical instead of spherical outflows and different velocity, density and pressure profiles with respect to those found here and could be investigated in future studies. Our assumption is that the hot wind subtends a solid angle of $4\pi$, while assuming a smaller angle would increase the wind densities, potentially leading to values closer to the X-ray observations (see Section~\ref{disc:xraydata}). Previous toy models \citep{diteodoro18,lockman20} have shown that the kinematics of our population of \hi\ high-velocity clouds can be described by a biconical wind with an opening angle of about $140^{\circ}$. This implies that the hot wind of our physical models needs to subtend at least such an angle in order to be responsible for the cloud entrainment. Reducing the opening angle to this value leads to an increase in normalization of the density (and pressure) profile of a factor $1.5$, which would not change any of the findings of this paper.

Finally, we recall that in our model the hot wind is expanding in vacuum, while in reality the outflow is most likely interacting with the pre-existing hot gas components \citep[e.g.,][]{miller16,locatelli24}. This impact would likely modify the wind dynamics and properties. \cite{sarkar24} roughly estimates the shock radius due to a galactic outflow expanding for a few megayears in the CGM, using the same expressions that are classically used to calculate the shock radius of a stellar wind on the interstellar medium \citep[e.g.,][]{weaver77}, to be about 2-3 kpc. Within this radius, one can expect the solution of the wind to be consistent with the vacuum-expanding one, which justifies our approach, given that all our findings are focused within roughly 2 kpc from the Galactic disk. Future works, trying to model the multiphase MW wind out to larger distances, should more thoroughly treat the impact between the outflow and the pre-existing CGM.

\subsection{Uncertainties in the mixing between gas phases}\label{disc:MassExchange}
One of the fundamental features of our model is the mixing of mass and momentum between the hot wind and the cold clouds due to turbulent 
mixing between them. An important approximation that we have adopted throughout this work is that the hot wind, even though it is continuously exchanging mass and momentum with the cold phase, maintains its properties fixed during the passage of the clouds and it is practically not affected by them. Here, we discuss how this simplification might influence our final findings.

\begin{figure}
    \centering
    \includegraphics[clip, trim={0cm 0cm 0cm 0cm}, width=\linewidth]{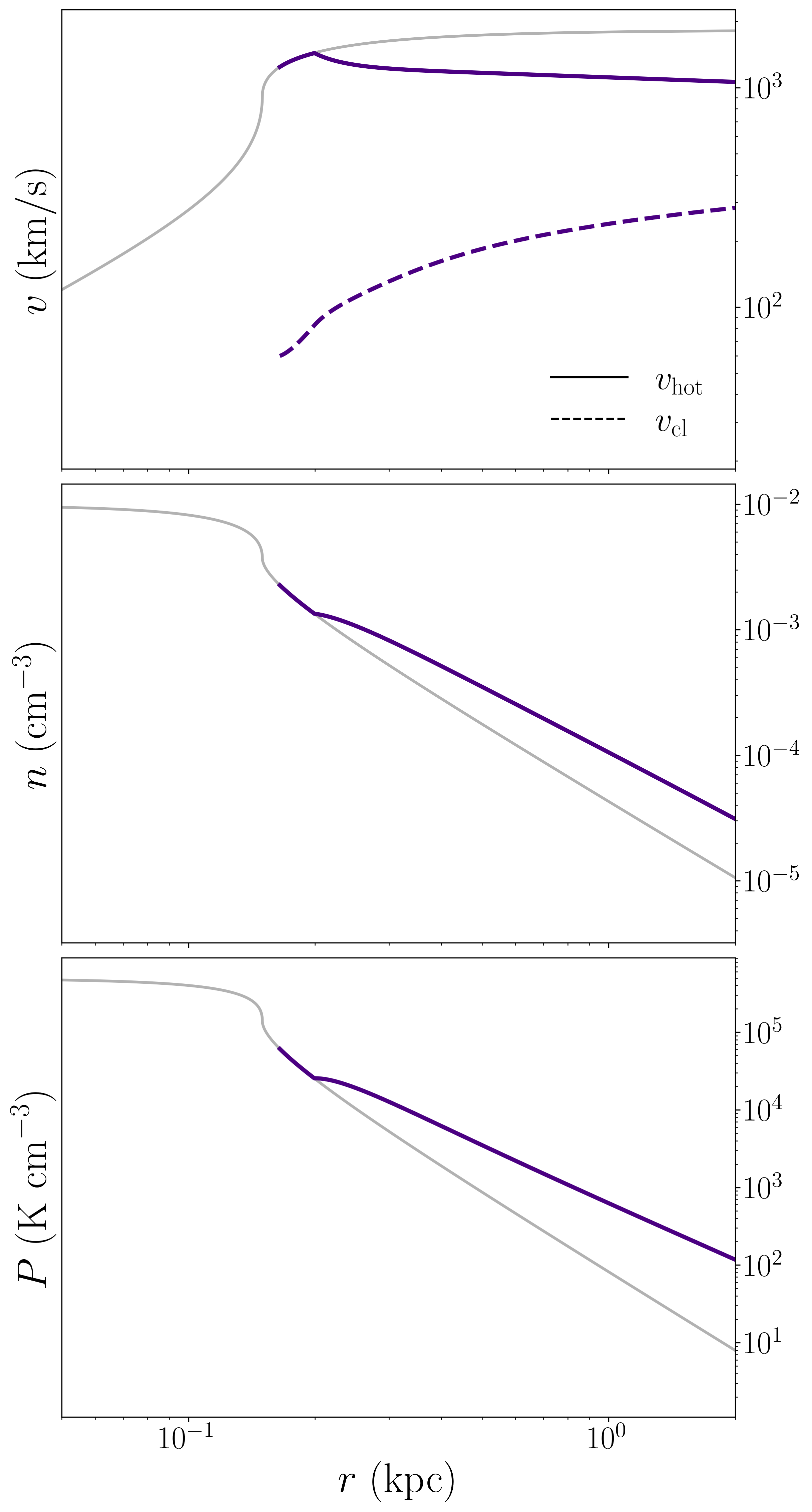}
    \caption{Velocity (top), density (center), and pressure (bottom) profiles of a hot-wind model (solid curves) consistent with the best-fit models found in Section~\ref{ModelResults}. In gray we report a model where the exchange of mass and momentum with the clouds has no effect on the hot phase, while in purple we show the results from the same model, but taking into account the full framework of \citetalias{fielding22}. The dashed purple curve in the top panel shows the corresponding radial velocity of the cold clouds for such a model.}
    \label{fig:FBmultiphase}
\end{figure}

We consider a radial multiphase model built entirely according to the full framework of \citetalias{fielding22} (see the equations of their Figure 3) and assuming initial conditions that are consistent with our best-fit models of Section~\ref{ModelResults}. Note that in here, the gravitational force influencing the cloud motion is only given by a logarithmic potential with circular velocity equal to $150$ \kms, which represents the main difference with our models, where we instead consider a full axisymmetric representation of the MW potential (see Section~\ref{sec:modelCold}). For this model, we used, for the hot phase, mass and energy loading factors equal to  0.15 and 0.5, respectively, and an injection radius equal to $150$ pc; for the cold phase, we placed just outside the injection radius clouds with initial masses of $100\ M_{\odot}$ and initial kick velocities of $60$ \kms,  with a total cold gas mass loading factor equal to 0.2, and we assumed $f_{\rm mix}=1/3$. All the values above are in agreement with our fiducial best-fit models (see Figures~\ref{fig:cornerMultiph} and \ref{fig:Lfactors}). The hot wind properties without any exchange of mass and momentum from the clouds (as in the rest of this work) are shown by the gray curves in Figure~\ref{fig:FBmultiphase}. The solid purple curves show instead the properties of the hot gas once the exchange of mass and momentum is taken into account: one can see that the velocities decrease by less than a factor of two and they are still at about 1000 \kms, and that the profiles of density and pressure are flatter with respect to the previous case, in agreement also with the results of hydrodynamical simulations \citep[e.g.,][]{schneider20}. In the top panel, the dashed curve shows the predicted cloud radial velocity, showing that also in this case the clouds are significantly accelerated by the wind to velocities of $300$ \kms\ at a radius of $2$ kpc. This is qualitatively in agreement (also considering the different gravitational potential of the two approaches) with the cloud velocities predicted by our fiducial models, implying that, despite the changes in the hot gas properties, the main findings of this work would not be altered.

 As we have shown above and in Section~\ref{ModelResults}, both the pressure exerted by the hot phase through the drag force and the accretion of high-velocity material are crucial to accelerating the \hi\ clouds and reproducing the observed kinematics. In particular, the dimensionless parameter $\xi$ determines whether the clouds experience an overall growth due to the gas mixing or whether they instead lose mass to the hot wind. We find that in our best-fit models $\xi<1$ everywhere, which is consistent with the fact that our clouds are evaporating and becoming less massive with time (see Section~\ref{ModelResults}). Another important parameter is $f_{\rm{mix}}$, which determines the normalization of the mass loss/gain rate (equation~\ref{eq:masstot}) and that we left as a free parameter in our modeling. As we have seen in Section~\ref{sec:modelCold}, this factor determines in practice the efficiency of the mixing between the cold and the hot phases of the outflow. Previous cloud-crushing simulations \citep[e.g.,][]{scannapieco15} have shown that clouds tend to be disrupted on timescales of $N_{\rm{cc}}$ cloud crushing times, where $\dot{m}_{\text{cl,loss}} \sim 1/N_{\rm{cc}}\ m_{\rm{cl}}/t_{\rm{cc}}$. Comparing this with our equation~\eqref{eq:massloss} implies that $N_{\rm{cc}}=10/(3 f_{\rm{mix}})$, so that for our best-fit models $N_{\rm{cc}}\sim11$. Interestingly, this value is roughly consistent with the results of previous works \citep[e.g.,][]{scannapieco15,schneider17}, where $N_{\rm{cc}}\gtrsim10$ in the supersonic region (which is our region of interest, given that the cold clouds are injected outside of the sonic radius) and is instead lower in the subsonic region.

We stress that the above parametrization of $N_{\rm{cc}}$, as well as the typical value of $f_{\rm{mix}}$, is estimated from the results of previous hydrodynamical simulations. However, these simulations do not include, among others, physical effects such as cosmic rays \citep[e.g.,][]{ruszkowski23,armillotta24} and thermal conduction \citep[see, e.g.,][]{bruggen16,armillotta16,afruni23sim} and its potential interplay with magnetic fields \citep[e.g.,][]{kooij21} nor variation in the hot background gas \citep[e.g.,][]{dutta25}: all these effects could potentially influence the amount of mixing between the different gas phases and therefore the cloud entrainment and survival. Our method of directly comparing analytical models with data can be used as a complementary way to infer the parameters (such as $f_{\rm{mix}}$) of subgrid models (e.g.,\ \citetalias{fielding22}; \citealt{huang20subgrid}) that can later be applied to cosmological simulations \citep[e.g.,][]{huang2022cosmo,weinberger23,smith24} and can therefore be crucial for the future treatment of gas mixing in galaxy formation.

\section{Summary and conclusions}\label{conclusions}
In this work, we have created semi-analytical models of a multiphase Galactic wind, powered by SN feedback, with the aim of interpreting the observational data of \cite{cluregriffiths13} and \cite{diteodoro18}, who through 21-cm emission found a population of \hi\ high-velocity clouds close to the Galactic center. In our parametric models, which we built using {\sc galpy} \citep{bovy15} and the framework from \citet{fielding22}, the hot gas phase tends to entrain and accelerate the cold clouds against the MW gravitational potential, through the effects of the drag force and of the exchange of mass (and therefore momentum) between the two phases. We infer the values of the model free parameters by performing a Bayesian analysis between the model outputs and the data. 

Our main findings are the following:

\begin{enumerate}[itemsep=1pt]
    \item We find a well constrained class of multiphase winds, driven by the star formation in the CMZ of the MW, that can successfully reproduce the locations, velocities and masses of the observed \hi\ cloud population;
    \item The hot gas phase of our best-fit models is well justified from a theoretical point of view, given that we infer mass and energy loading factors in very good agreement with expectations from recent high-resolution hydrodynamical simulations. Our predicted hot winds are fainter and hotter compared to current X-ray constraints: this could either mean that we are modeling a different component to that detected in the observations, or that larger mass injection rates     are needed;
    \item The cold clouds have masses and (roughly) radii in agreement with the observational estimates and are accelerated by the interactions with the hot wind to velocities of $300-400$ \kms. We also find that, in order to reproduce the observations, the clouds need to overall lose mass to the hot component, with the least massive clouds having lost more than $70\%$ of their initial mass after $10$ Myr.
    
\end{enumerate}

This work therefore provides a physically motivated and self-consistent model that is able to reproduce the kinematics of this population of \hi\ high-velocity clouds. With this analysis, we have found that SN feedback (even though we cannot exclude the influence of the central AGN) could be a natural driver of the multiphase outflow at the center of our Galaxy, and it is therefore a crucial process to take into account for the creation of the Fermi Bubbles.

\begin{acknowledgements}
    We thank Kartic Sarkar, Gabriele Ponti, Rongmon Bordoloi, Drummond Fielding, Alankar Dutta, Gabriele Pezzulli and Filippo Fraternali for helpful discussions that have contributed in developing some of the ideas presented in this work. AA and EDT were supported by the European Research Council (ERC) under grant agreement no.\ 101040751. LA was supported by the INAF Astrophysical fellowship initiative.
\end{acknowledgements}

\bibliographystyle{aa}
\bibliography{references}

\begin{appendix}
\section{Ballistic model}\label{sec:ballistic}
In this appendix, we investigate whether a ballistic model would be able to reproduce the kinematic data described in Section~\ref{observations} and shown in Figure~\ref{fig:data}, to assess whether or not a hot entraining wind is needed to explain the observed kinematic pattern. For this purpose, we modeled the cloud orbits using {\sc galpy} as in Section~\ref{sec:modelCold}, but including only the MW gravitational potential and no other additional force due to the influence of the hot wind. We assumed that all the cold clouds start within the injection radius ($r_{\rm{inj}}$, which in this case obviously refers only to the cold gas), with initial $(R_{\rm{in}},z_{\rm{in}})$ for each orbit chosen exactly as in Section~\ref{sec:modelCold}. As for the initial cloud velocities, we assumed an initial kick velocity, $v_{\rm{kick}}$, as in our fiducial models. Differently from Section~\ref{sec:modelCold}, we did not assume that $v_{\rm{kick}}$ is entirely vertical (that was previously justified by the fact that the influence of the hot wind would dominate the cloud dynamics), but we assumed an angle, $\theta_{\rm{max}}$, that determines the extent of the biconical outflow aperture. For each orbit, we indeed selected a random $\theta$ uniformly sampled between $0$ and $\theta_{\rm{max}}$. We additionally chose for each orbit a random angle $\phi$ between $0$ and $2\pi$, such that \citep[see][]{afruni21}:
\begin{ceqn}
\begin{equation}\label{eq:invel}\begin{aligned}
v_{\rm{kick,R}}=v_{\rm{kick}}\sin{\theta}\cos{\phi}\ ,\\
v_{\rm{kick,T}}=v_{\rm{circ}}+v_{\rm{kick}}\sin{\theta}\sin{\phi}\ ,\\
v_{\rm{kick,z}}=v_{\rm{kick}}\cos{\theta}\ .
\end{aligned}
\end{equation}
\end{ceqn}

In this framework, the free parameters of our models are therefore $r_{\rm{inj}}$, $\theta_{\rm{max}}$ and $v_{\rm{kick}}$. For each choice of these free parameters we modeled 20 different orbits for $10$ Myr, with initial conditions as explained above, and we compared the model clouds populating these orbits (created exactly in the same way as in Section~\ref{sec:population}) with the observed \hi\ components, using the likelihood defined by equation~\eqref{eq:likelihood}. In particular, we ran a Bayesian analysis with {\sc dynesty}, with the following priors: a flat prior on the injection radius $0.05<r_{\rm{inj}}/\rm{kpc}<0.3$ (in order to be consistent with the CMZ extent, see also Section~\ref{sec:likelihood}); a flat prior in $\theta_{\rm{max}}$ between 0 (vertical outflow) and $\pi/2$ (isotropic outflow); a flat prior in the kick velocity $0.5<v_{\rm{kick,100}}<5$.

\begin{figure}
    \centering
    \includegraphics[clip, trim={0cm 0cm 0cm 0cm}, width=\columnwidth]{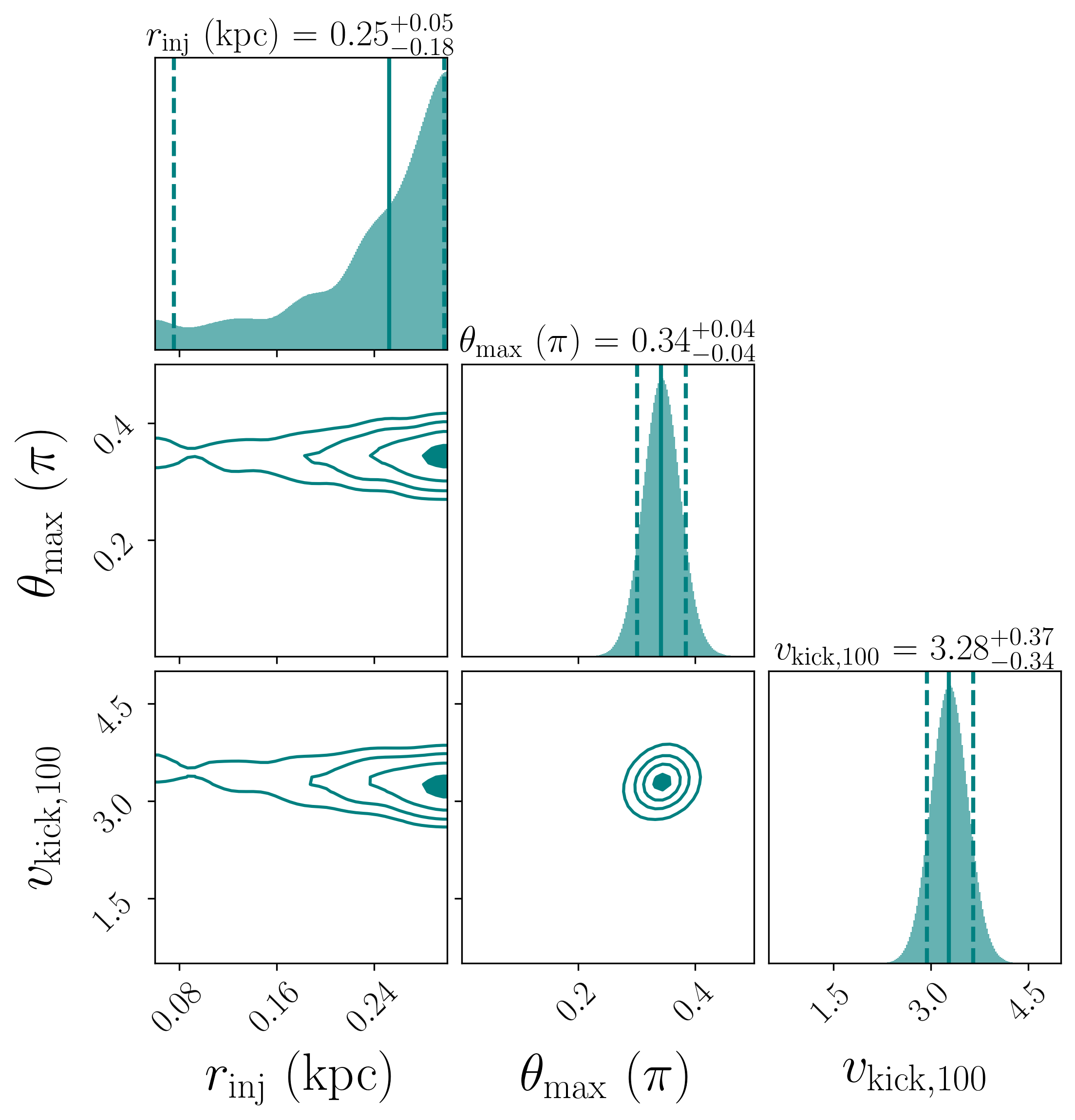}
    \includegraphics[clip, trim={0cm 0cm 2cm 0cm}, width=\columnwidth]{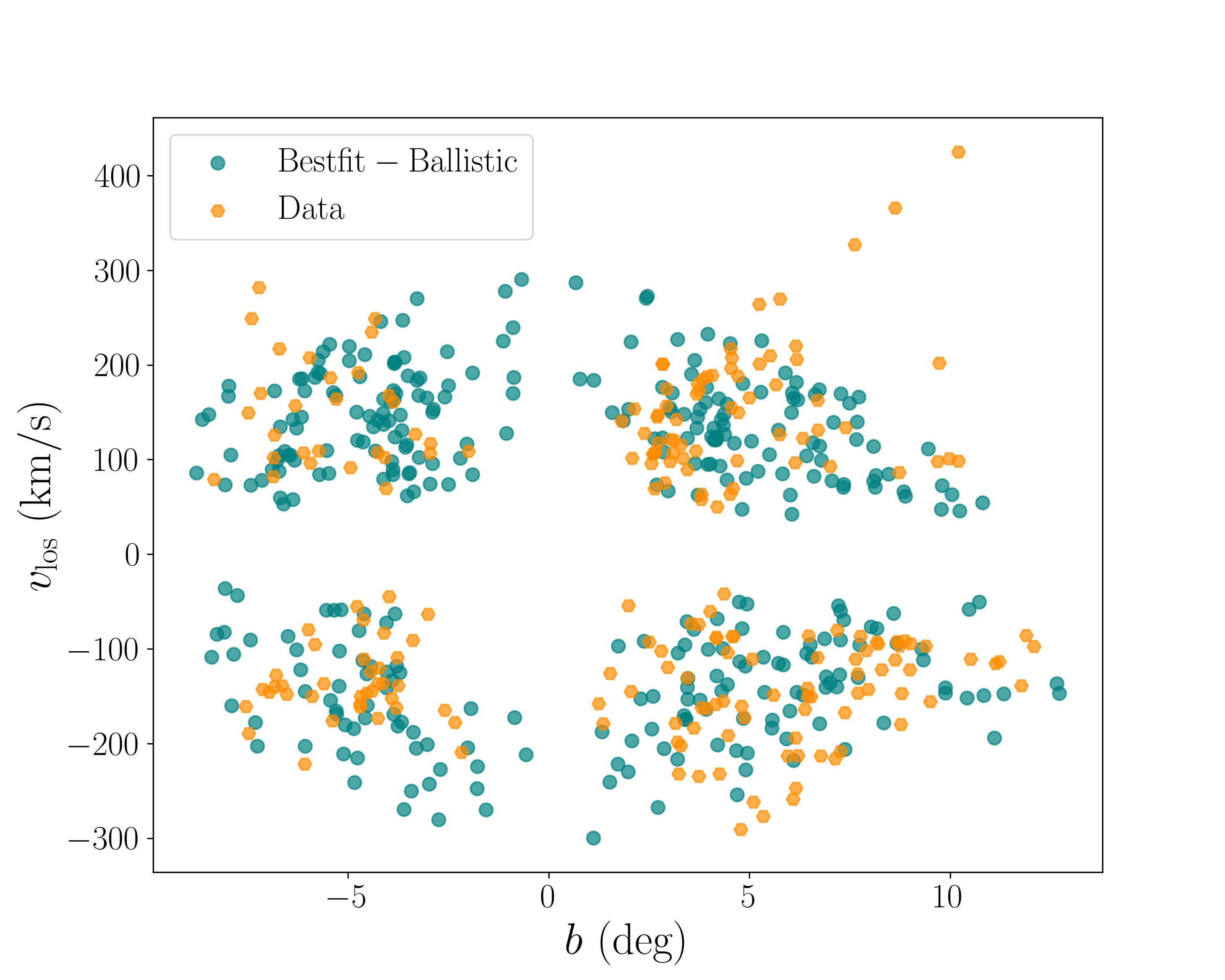}
    \caption{Main results of the ballistic models presented in Appendix~\ref{sec:ballistic}. Top: corner plot showing the posterior distributions of the three free parameters; the vertical lines mark the positions of the median values of the posteriors and of the $2-\sigma$ uncertainties. Bottom: comparison between the outputs of the best-fit models (teal) and the observations (orange).}
    \label{fig:cornerPlotbal}
\end{figure}

We show the results of this analysis in Figure~\ref{fig:cornerPlotbal}. In the top panel, we report the corner plot of the posterior distributions of the three free parameters: one can see that in order to reproduce the data the models need large injection radii ($r_{\rm{inj}}\gtrsim250$ pc), relatively large outflow apertures ($\theta_{\rm{max}}\sim \pi/3$) and large initial kick velocities ($v_{\rm{kick}}\sim330$ \kms). In particular, both the outflow velocity and aperture are very well constrained and exhibit narrow posterior distributions. In the bottom panel of Figure~\ref{fig:cornerPlotbal} we show instead how these best-fit models perform in reproducing the data. Here, we report the cloud velocity as a function of the Galactic latitude, with the data in orange and the best-fit models in teal (similar to Figure~\ref{fig:comparison_new}). One can immediately see that, contrary to our fiducial multiphase models, a ballistic model is not able to reproduce the observed kinematic pattern of acceleration. The models show indeed, as expected, the opposite behavior of deceleration, given that the gravitational force tends to slow down the clouds that are initially ejected with very high velocities. This comparison therefore demonstrates how the motion of the \hi\ clouds detected in the vicinity of the Galactic center cannot be explained only by gravity.

Note that taking into account the impact of the cold outflow with a static or slowly rotating hot CGM \citep[e.g.,][see also Section~\ref{disc:WindAssumptions}]{miller15,hodges16,locatelli24}, similar to the models implemented in \cite{afruni21,afruni22}, would make the model outputs even more inconsistent with the data, as it would increase the amount of deceleration of the clouds. We therefore conclude that a fast-moving hot wind is needed to reproduce the observed kinematic pattern of the \hi\ high-velocity clouds.

\section{Massive hot wind model}\label{sec:massiveWind}
In Section~\ref{disc:comparisonObs} we have discussed a model where we imposed the hot wind mass loading factor to be equal to 1, by adopting in our Bayesian analysis a narrow Gaussian prior centered in $\log \eta_{\rm{M}}=0$. This leads to a slower, denser, and cooler wind, in better agreement with the observational constraints from the X-ray chimneys data (see Figure~\ref{fig:HotgasProp}). Even though our preferred models have lower mass loading factors, more consistent with theoretical predictions (see discussion in Section~\ref{disc:xraydata}), we show here that if we considered a more massive wind model the observed \hi\ cloud kinematics would still be reproduced by our framework.

\begin{figure}
    \centering
    \includegraphics[clip, trim={0.26cm 0.23cm 0.2cm 0.2cm}, width=\columnwidth]{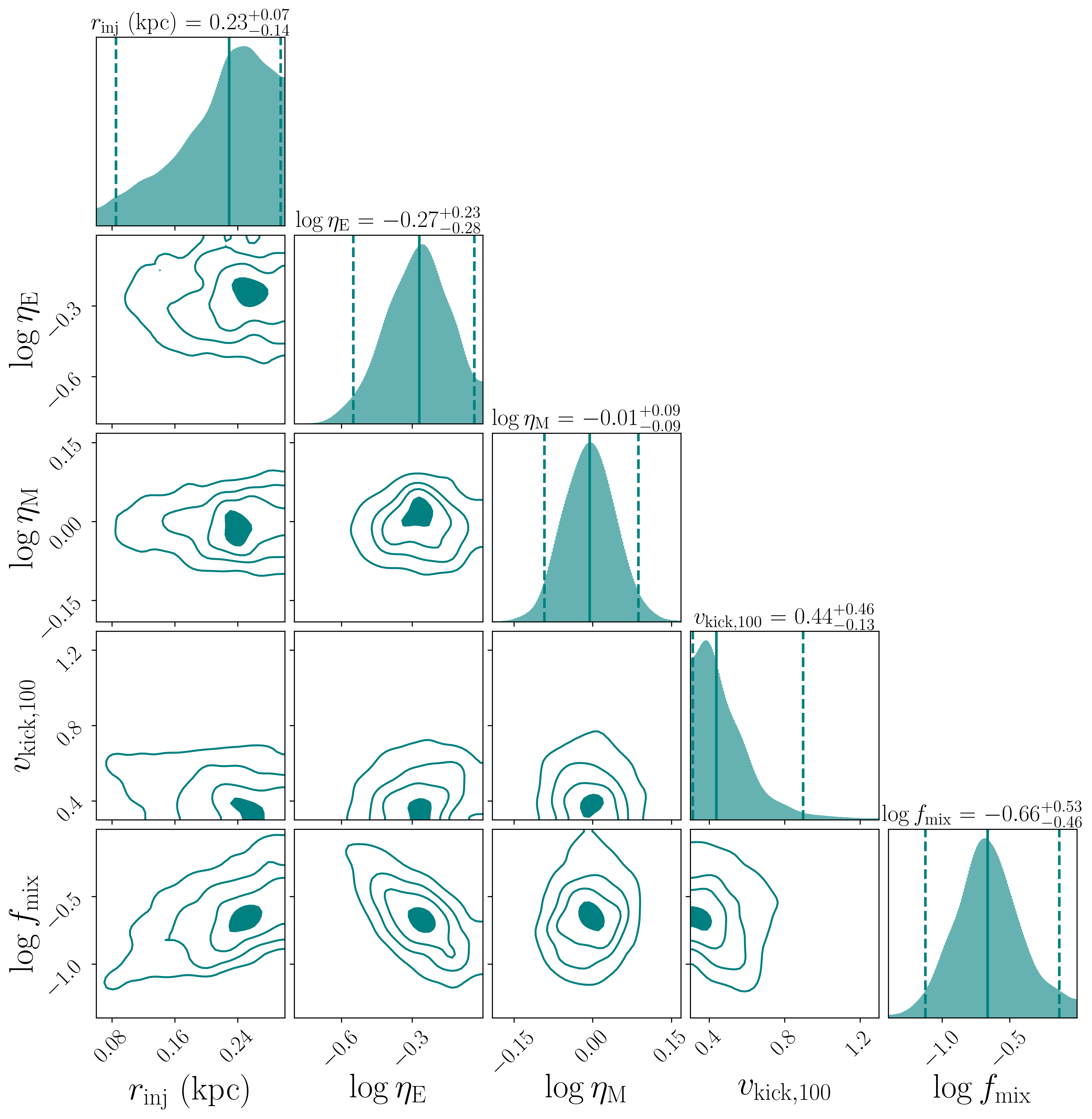}
    \caption{Corner plot showing the posterior distributions of the five free parameters of the model presented in Section~\ref{disc:xraydata} and Appendix~\ref{sec:massiveWind}, where a narrow Gaussian prior centered in $\log \eta_{\rm{M}}=0$ is adopted for the hot wind mass loading factor. As for our fiducial model (see Section~\ref{sec:likelihood}), we also adopted a Gaussian prior for the energy loading factor, centered in $\log \eta_{\rm{E}}=-0.4$. The vertical lines mark the positions of the 2.5, 50 (solid line) and 97.5 percentiles of the 1-dimensional posterior distributions.}
    \label{fig:corneretaM0}
\end{figure}

\begin{figure}
    \centering
    \includegraphics[clip, trim={0cm 0cm 0cm 0cm}, width=\columnwidth]{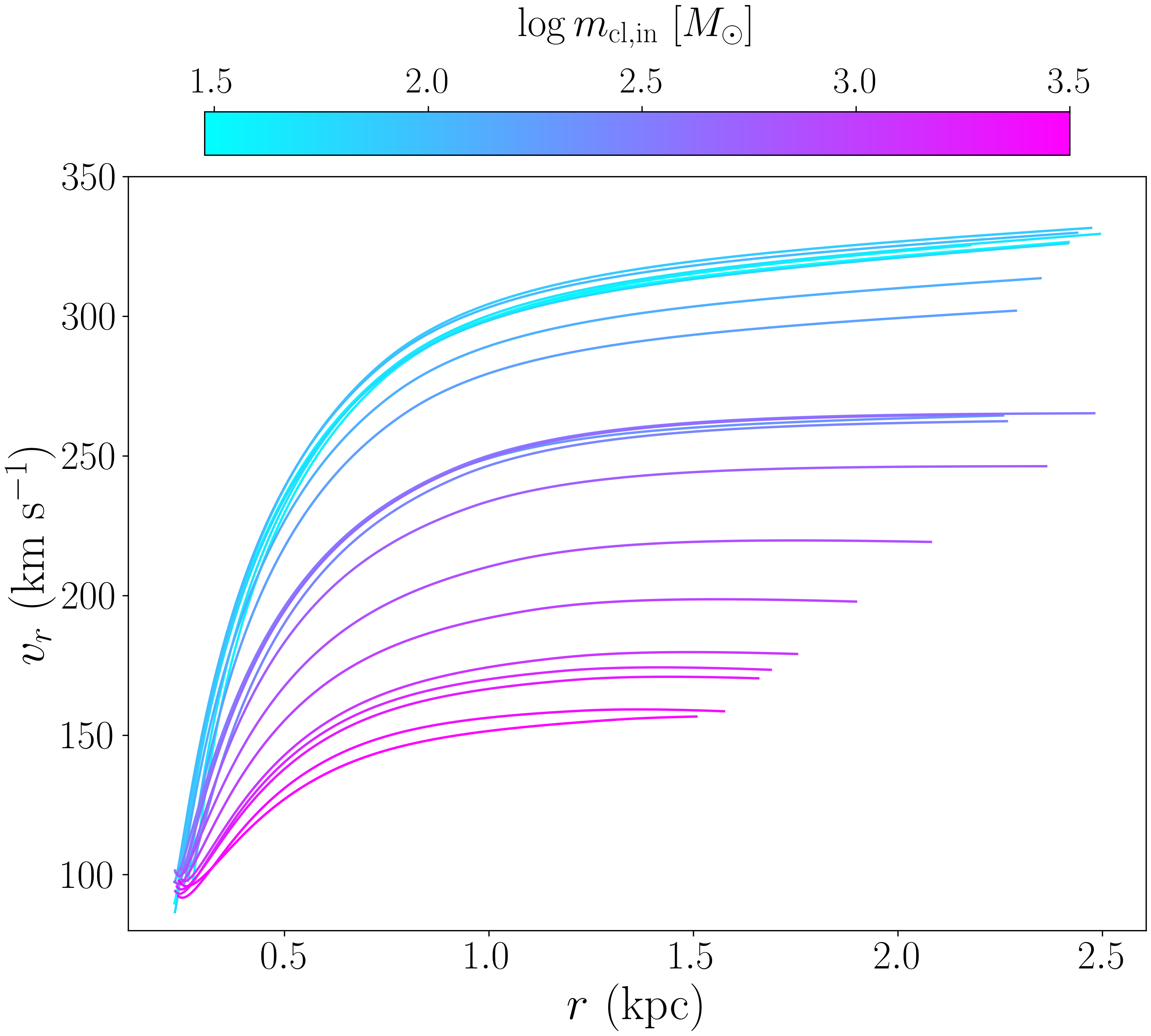}
    \includegraphics[clip, trim={0cm 0cm 0cm 0cm}, width=\columnwidth]{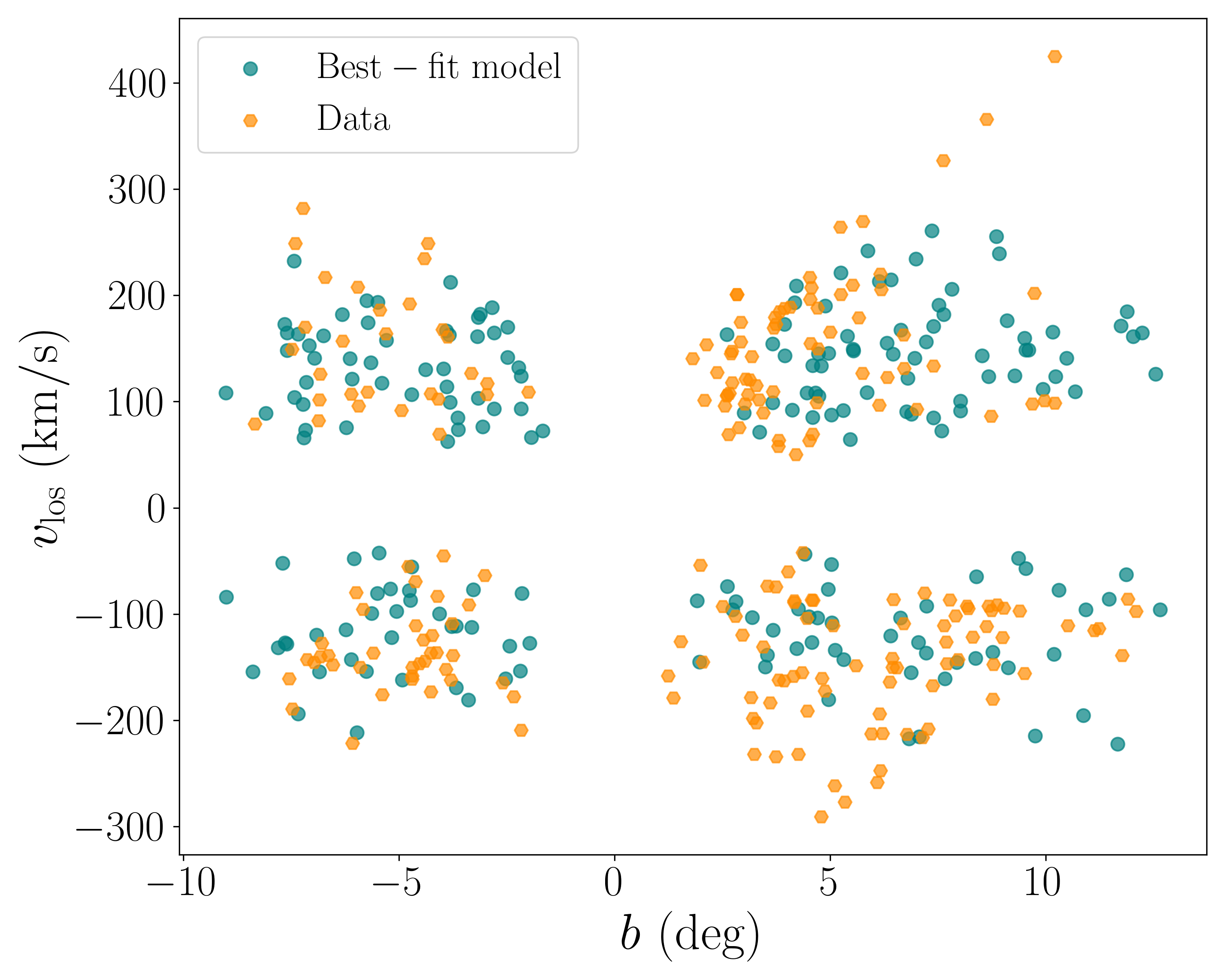}
    \caption{Outputs of the best-fit model assuming a massive hot wind (see Figure~\ref{fig:corneretaM0} and Appendix~\ref{sec:massiveWind}). Top: Radial velocities of 20 typical clouds as a function of the distance from the center, color-coded by the cloud initial mass. Bottom: Comparison between the outputs of the best-fit models (teal) and the observations (orange).}
    \label{fig:kinematics_etaM0}
\end{figure}

In the two panels of Figure~\ref{fig:kinematics_etaM0}, we show the outputs of a model where the values of the 5 parameters are fixed to the median values of the posterior distributions obtained with the Bayesian analysis (Figure~\ref{fig:corneretaM0}). In the top panel, we show the evolution of the total radial velocities (obtained by summing in quadrature the three velocity components) of 20 typical clouds in this model. Note how, as for our fiducial models, the clouds are accelerated by the interactions with the hot wind, with the lower mass clouds reaching higher velocities compared to the more massive ones. Compared to the scenario presented in Section~\ref{ModelResults}, where the clouds keep accelerating with increasing distance from the center, here they appear to reach a maximum velocity of about 300 \kms\ (for the lower mass clouds). This effect is likely due to the lower velocities of the hot wind in this model, as can be seen in Figure~\ref{fig:HotgasProp}. As a result (see bottom panel of Figure~\ref{fig:kinematics_etaM0}), this model does not reproduce the few clouds at high latitudes with velocities approaching $400$ \kms, which were instead recovered by our fiducial model. However, except for this small discrepancy, the cloud kinematic distribution is still consistent with the observed one. This leads us to conclude that a denser wind would still be able to overall explain the kinematics of this population of \hi\ high-velocity clouds close to the Galactic center.

\end{appendix}

\end{document}